\documentclass{aa}  

\usepackage{dcolumn}
\usepackage{graphicx}
\usepackage{changepage}
\usepackage{txfonts}
\usepackage{soul}
\usepackage{pifont}
\usepackage[breaklinks=true,colorlinks = true, linkcolor= blue, citecolor = blue, filecolor = blue, urlcolor = blue]{hyperref}
\begin{document}

   \title{Runaway blue main-sequence stars at high Galactic latitudes}

   \subtitle{Target selection with {\em Gaia} and spectroscopic identification}

   \author{Roberto Raddi
          \inst{1,2}\fnmsep\thanks{\email{roberto.raddi@upc.edu}}
          \and
          Andreas Irrgang\inst{2}
          \and
          Ulrich Heber\inst{2}
          \and
          David Schneider\inst{2}
          \and
          Simon Kreuzer\inst{2}
          }

   \institute{Universitat Polit\`ecnica de Catalunya, Departament de F\'isica, c/ Esteve Terrades 5, 08860 Castelldefels, Spain \and
   Dr.~Karl~Remeis-Observatory \& ECAP, Astronomical Institute,
Friedrich-Alexander University Erlangen-Nuremberg (FAU),
Sternwartstr.~7, 96049 Bamberg, Germany}

   \date{Received xx; accepted yyy}

 
  \abstract
   {The ESA {\em Gaia} mission is a remarkable tool for stellar population analysis through its accurate Hertzsprung-Russell diagram. Its precise astrometry has propelled detailed kinematic studies of the Milky Way and the identification of high-velocity outliers.}
   {Motivated by the historical identification of runaway main-sequence (MS) stars of early spectral type at  high Galactic latitudes, we test the capability of {\em Gaia} at identifying new such stars.}
   {We have selected $\approx 2300$ sources with {\em Gaia} magnitudes of $G_{\rm BP} - G_{\rm RP} \leq 0.05$, compatible with the colors of low-extinction MS stars earlier than mid-A spectral type, and obtained low-resolution optical spectroscopy for 48 such stars. By performing detailed photometric and spectroscopic analyses, we derive their atmospheric and physical parameters (effective temperature, surface gravity, radial velocity, interstellar reddening, spectrophotometric distance, mass, radius, luminosity, and age). The comparison between spectrophotometric and parallax-based distances enables us to disentangle the MS candidates from older blue horizontal branch (BHB) candidates. }
   {We identify 12 runaway MS candidates, with masses between 2 and 6\,M$_{\sun}$. Their trajectories are traced back to the Galactic disc to identify their most recent Galactic plane crossings and the corresponding flight times. All 12 candidates are ejected from the Galactic disc within 2 to 16.5\,kpc from the Galactic center and possess flight times that are shorter than their evolutionary ages, compatible with a runaway hypothesis.
   Three MS candidates have ejection velocities exceeding 450\,km\,s$^{-1}$, thus, appear to challenge the canonical ejection scenarios for late B-type stars. The fastest star of our sample also has a non-negligible Galactic escape probability if its MS nature can be confirmed. We identify 27 BHB candidates, and the two hottest stars in our sample are rare late O and early B type stars of low mass evolving towards the white dwarf cooling sequence. }
   {The combination of {\em Gaia} parallaxes and proper motions can lead to the efficient selection of runaway blue MS candidates up to 10\,kpc away from the Sun. 
 High resolution spectra are needed to confirm the MS status, via precise measurements of projected rotational velocities and chemical compositions. }


   \maketitle
%
%
\section{Introduction}

The presence of short-lived, early-type main sequence (MS) stars at high Galactic latitudes has long been a puzzle that, now more than ever, demonstrates the existence of violent astrophysical phenomena, which lead to the ejection of these stars from their birth places such as open clusters and OB associations. 

In the classical picture, Galactic runaway stars are accelerated through variations of the {\em binary supernova mechanism} \citep{blaauw1961} and {\em dynamical ejection} from open clusters \citep{poveda1967}. In the first case, the runaway star is ejected subsequently to the core-collapse supernova of a more massive companion; in the second case, binary interactions in massive clusters provide the ejection mechanism. Both phenomena are seen to take place early in the stellar lifetime, within the same regions of high-mass star formation \citep{hoogerwerf2001}. The predicted ejection velocity distribution of these mechanisms ranges from a few tens of km\,s$^{-1}$ up to 400--500\,km\,s$^{-1}$, strongly depending both on the mass of the runaway star and on the kind of interactions \citep{portegieszwart2000, gvaramadze2009, tauris2015, renzo2019,evans2020}.
The binary ejection mechanism is highly efficient at forming runaway stars if a ``birth kick'' is imparted as consequence of asymmetric supernova explosions \citep[see discussions by][]{renzo2019,evans2020}.
For the dynamical ejection scenario, the birth rate of runaway stars is seen to strongly correlate with increasing cluster density and stellar mass \citep{perets2012, oh2016}. The fastest runaway stars are expected to form mostly via the binary supernova mechanism and to reach no further away than a few tens of kilo-parsecs above the Galactic plane, implying that just a negligible fraction of them might escape from the Milky Way. 

The literature is rich with works focusing on the identification and classification of runaway stars. Notably, \citet{silva2011} produced a large systematic study of previously known runaway blue stars, arguing that their ejection velocity distribution is compatible with the binary supernova ejection scenario, although suggesting that a small contribution could also arise from the dynamical ejection from open clusters. The assortment of runaway early-type MS stars is not just populated by former Galactic disc members. In fact, there are extra-galactic visitors like \object{HE\,0437$-$5439} 
(also known as HVS\,3) that seems to have been ejected from the Large Magellanic Cloud\\ \citep[][]{edelmann2005, irrgang2018b, erkal2019a}. 
In addition, other more energetic phenomena have also been suggested to operate in the Milky Way. One of them is the {\em Hills mechanism} \citep[][]{hills1988}, 
which could be responsible for the formation of the {\em hyper-velocity} stars through the interaction of binaries with the supermassive black hole at the center of our Galaxy. With ejection velocities in the range  $1000$\,km\,s$^{-1}$, these stars could also escape the Milky Way \citep[see][for a review]{brown2015}.

Wider samples of runaway stars have recently become more accessible thanks to the ESA {\em Gaia} astrometric mission \citep{gaia}. Following the promising search for runaway stars in the first data release of {\em Gaia} \citep{maiz-appelaniz2018b}, the second data release \citep[DR2;][]{gaia-dr2} has been the real game changer, due to the high precision of its space-based proper motions, parallaxes, broad-band photometry, and radial velocities \citep[e.g.][]{boubert2018,brown2018,irrgang2018b,shen2018,hattori2018, raddi2018, raddi2019, koposov2020, kreuzer2020}.

In this manuscript, we present a selection of blue stars at high Galactic latitudes, guided by the {\em Gaia} Hertzsprung-Russell diagram (Sect.~\ref{sec:selection}), with the goal of testing the prospects for the identification of blue MS runaway candidates in our Galaxy. We obtained low-resolution optical spectra for 48 stars (Sect.~\ref{sec:observations}), which enable us to determine effective temperatures, surface gravities, and radial velocities. We classify 12 likely MS stars, 27 blue horizontal branch (BHB) stars, and two very hot evolved stars via the agreement between spectrophotometric and parallax-based distances (Sect.~\ref{sec:analysis}). We determine the Galactic orbits of the MS candidates (Sect.~\ref{sec:orbits}), and discuss their key properties in the context of runaway MS stars (Sect.~\ref{sec:discussion}).  The results of this work suggest future improvements for the identification of runaway stars by exploiting the synergy between {\em Gaia} and optical spectroscopy.


\section{Target selection}
\label{sec:selection}
   \begin{figure}
   \centering
   \includegraphics{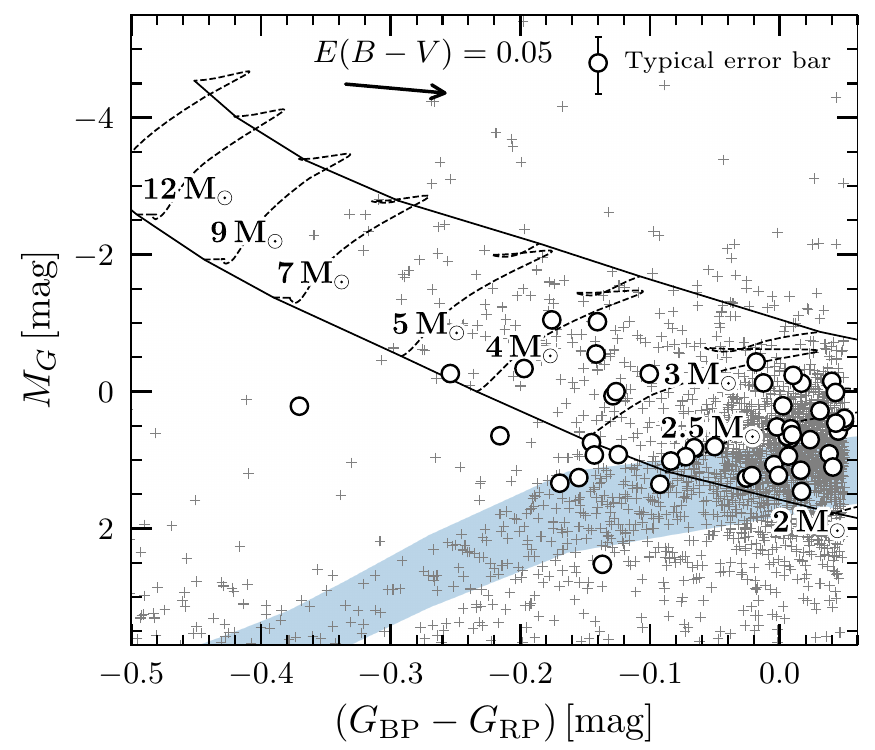}

   \caption{{\em Gaia} Hertzsprung-Russell diagram of candidate blue stars at high Galactic latitudes (grey cross symbols). The observed targets are  shown as large circles. Solid curves represent the zero-age and terminal-age MS, while the evolutionary tracks for representative initial masses are  plotted  as dashed  curves  \citep[][]{choi2016}. The light-blue strip represents the  horizontal branch \citep{dorman1992}. The reddening vector and typical error bars of the observed targets are displayed at the  top edge of the figure.}
              \label{fig:hr}
    \end{figure}
   \begin{figure}
   \centering
   \includegraphics{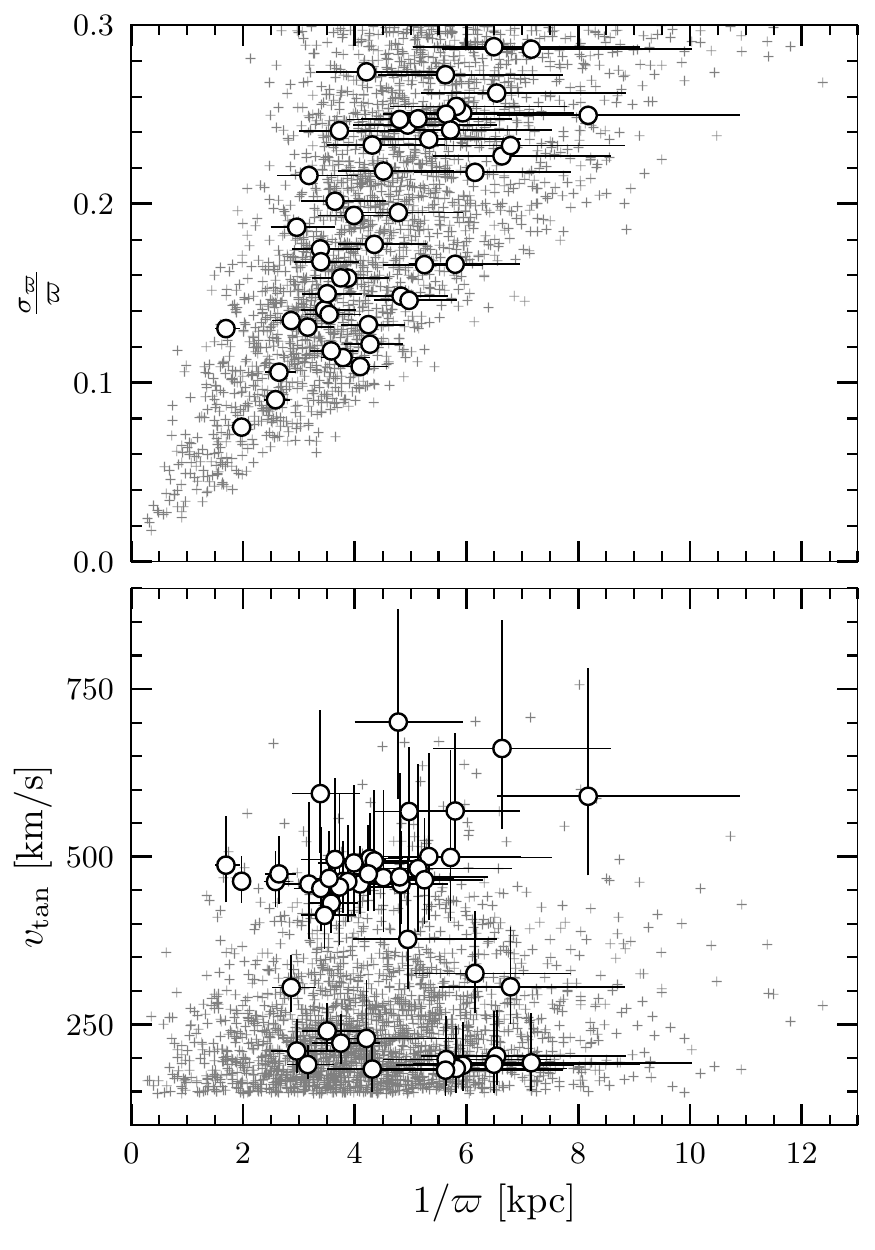}
   \caption{The relative parallax uncertainties and tangential velocities of 
   early-type star candidates (cross symbols) and observed targets
   (circles with error bars) are plotted against the parallax-based distance
   in the top and bottom panels, respectively.}
    \label{fig:plx-vel}
    \end{figure}
We queried the {\em Gaia} DR2 in order to identify candidate MS runaway stars with masses above 2\,M$_{\sun}$ at high Galactic latitudes ($|b| \geq 15^{\circ}$). We required that parallaxes are measured with a precision of better than 30\%, and we applied color and absolute magnitude cuts of $G_{\rm BP} - G_{\rm RP} \leq 0.05$\,mag and $M_{G} = G + 5 - 5\log{(1000/\varpi)} \leq 3.7$\,mag, respectively, where $G$, $G_{\rm BP}$, and $G_{\rm RP}$ are the {\em  Gaia} magnitudes, and $\varpi$ is the parallax given in mas\,yr$^{-1}$. We filtered the  {\em Gaia} photometry and astrometry,  requiring high quality data as prescribed by \citet{lindegren2018} and the \citet{gaia-hr}. In addition, we requested that our selection had re-normalized unit weight error (ruwe) smaller than 1.4 \citep{lindegren2018-b}. This procedure delivered 10\,342 objects.  The {\sc adql} code used to perform our query is given in Appendix~\ref{a:one}. 

We note that the $G$ magnitude distribution of the selected candidates is double peaked at $\approx 8.5$ and 14.5\,mag, and that apparently brighter objects present relatively small tangential velocities in a heliocentric reference frame. In order to reduce contamination from objects that are less-likely runaway stars, we applied an additional cut defined via the tangential velocity  as:  
\begin{equation}
\varv_{\rm tan} = 4.74 \times \frac{\sqrt{{\mu^{*}_{\alpha}}^{2} + \mu_{\delta}^{2}}}{\varpi} \geq 150\,\mathrm{km}\,\mathrm{s}^{-1};
\label{eq:vtan}
\end{equation}
where $\mu^{*}_{\alpha} = \mu_{\alpha} \cos{\delta}$ and $\mu_{\delta}$ are the proper motion components along the right ascension and declination directions, respectively. This velocity cut delivers 2392 objects that are displayed in Fig.~\ref{fig:hr}, along with the 48 stars for which we have collected follow-up spectroscopy.

Focusing on blue stars, our selection does not overlap with previous work \citep[e.g.][]{hattori2018},
which targeted cool stars for which six-dimensional data were delivered by {\em Gaia} DR2, including radial velocities that are not available for stars bluer than early-F spectral type, but also have their pitfalls \citep{2019MNRAS.486.2618B}. Instead, we note that towards the lower edge of Fig.~\ref{fig:hr}, i.e. the faint absolute magnitude cut, our candidates overlap with the hot subluminous star candidates identified by \citet{geier2019}, which we avoided as much as possible in our spectroscopic follow-up. In addition, we note that our target selection of blue stars at high Galactic latitudes would contain a large fraction of BHB stars and field blue stragglers \citep{2010AJ....139...59B} out of which we aim to disentangle MS runaway candidates.

\section{Observations}
\label{sec:observations}

The time allocation for our follow-up observations consisted of five nights at the New Technology Telescope (NTT)  on 2019 August 15--20, with the ESO Faint Object Spectrograph and Camera \citep[EFOSC2;][]{buzzoni1984}. 

\begin{figure*}
\centering
\includegraphics[width=17.5cm]{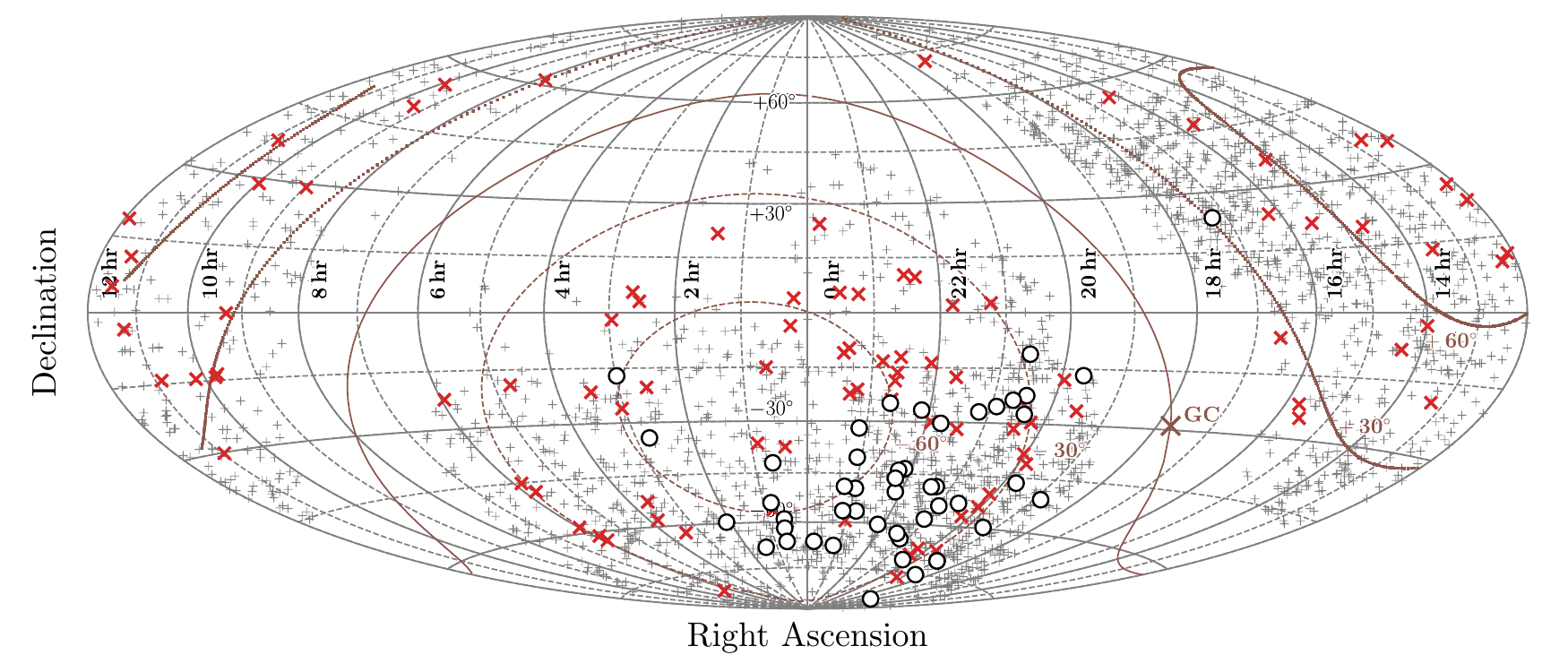}
\caption{On-sky distribution of {\em Gaia}-selected candidate blue stars at high Galactic latitude (grey '+' symbols). The 48 observed targets (circles) and known MS runaway stars \citep[red 'x' symbols;][]{silva2011} are also shown.}
\label{fig:sky}
\end{figure*}

Based on the on-sky visibility of suitable targets, we defined our priority list according to two empirical criteria. First, we favored objects having colors and absolute magnitudes that place them in the proximity of the unreddened MS (Fig.~\ref{fig:hr}). Second, we covered a wide range of tangential velocities so not to solely favor stars with the highest apparent motion, as this quantity is negatively affected by larger parallax uncertainties (Fig.~\ref{fig:plx-vel}). Due to the time allocation in August, we observed 48 stars with right ascension between 16.5\,h and 3\,h. Figure~\ref{fig:sky} shows the spatial distribution of all the identified candidates and observed targets, which are compared to known runaway stars \citep{silva2011}. The relevant {\em Gaia} data for the observed targets are listed in Table~\ref{t:phot}.

All the stars were observed with grism \#7 and $2 \times 2$ CCD binning, taking two exposures that ranged between 5--30\,min, followed by one He-Ar arc exposure at the same position of the observed star. We typically observed three white dwarf flux standards per night (GD\,50, LAWD\,74, and LTT\,7987). Our observations were undertaken near to full moon and the weather conditions were generally stable, frequently reaching sub-arcsec seeing; the sole exception was the first night, when thick clouds disrupted the second half of the night. 

We used standard {\sc iraf} \citep{iraf1986,iraf1993} procedures for the reduction of long-slit spectra \citep{iraf1992,iraf1997}, which we  bias subtracted, flat-field corrected, and wavelength and flux calibrated.

The observing setup delivered a useful spectral coverage between 3650$-$5050\,\AA, with a dispersion of $1.9$\,\AA/pixel (that translates to $\approx 120$\,km\,s$^{-1}$ at H$\beta$). Because the point-spread-function of the spectral trace was smeared due to tracking limitations, the resolving power is strictly determined by the slit width. We measured a spectral resolution of $\Delta \lambda = 6.5$\,\AA\ from the full width at half maximum of the He-Ar lines of the calibration lamp. The residuals of the wavelength solution have a root-mean-square of 0.2\,\AA. Given that no strong emission line from the night sky is detectable within the wavelength range of our spectra, we assessed the accuracy of the wavelength calibration by measuring the radial velocities of the white dwarf standards GD\,50 and LTT\,7987. Following the line-fitting technique adopted by \citet{napiwotzki2019}, we used their analysis software, {\sc fitsb2}, to measure the radial velocity, $\varv_{\rm rad}$, of the white dwarf standards by iteratively combining synthetic spectra \citep{koester2010} with Gaussian and/or Lorentzian functions.  From this procedure, we estimated an average systematic uncertainty of $\pm 25$\,km\,s$^{-1}$ with respect to the values published by \citet{napiwotzki2019}, which we will add in quadrature to that obtained for the radial measurements of our science targets.
\begin{table*}
\caption{{\em Gaia} DR2 data for the observed 48 targets. }
\centering
\scriptsize
\begin{tabular}{@{}ccc D{o}{\,\pm\,}{4} D{o}{\,\pm\,}{4} cccccc@{}}
\hline
Gaia DR2 & $\alpha$    & $\delta$   & \multicolumn{1}{c}{$\mu^{*}_{\alpha}$} & \multicolumn{1}{c}{$\mu_{\delta}$}  & $\varpi$ & $G$   & $G_{\rm BP}$ & $G_{\rm RP}$ & Short name	& Observing date\\
         & (hms) & (dms) & \multicolumn{1}{c}{(mas\,yr$^{-1}$)}    & \multicolumn{1}{c}{(mas\,yr$^{-1}$)} & (mas)	 & (mag) & (mag)       & (mag)       & (hhmm$\pm$ddmm) & (yyyy-mm-dd)\\
\hline
4907042024497066752 & 00:35:19.19 & $-$59:00:22.35 & 12.35o0.05 & -8.92o0.04 & $0.12\pm0.03$ & 14.47 & 14.46 & 14.44 & 0035$-$5900 & 2019-08-16\\
4901793883699014912 & 00:37:24.77 & $-$61:54:55.17 & 4.29o0.07 & 5.15o0.06 & $0.17\pm0.04$ & 14.64 & 14.56 & 14.71 & 0037$-$6154 & 2019-08-20\\
4707068828231737856 & 00:38:16.76 & $-$66:12:17.46 & 21.22o0.05 & -12.31o0.05 & $0.23\pm0.03$ & 13.86 & 13.85 & 13.84 & 0038$-$6612 & 2019-08-16\\
4992937934442887296 & 00:39:20.74 & $-$42:01:22.10 & 20.70o0.04 & -12.06o0.05 & $0.23\pm0.04$ & 14.04 & 13.99 & 14.05 & 0039$-$4201 & 2019-08-16\\
4921153294167369856 & 00:49:49.34 & $-$53:56:30.65 & 20.97o0.04 & -10.96o0.04 & $0.24\pm0.03$ & 13.79 & 13.77 & 13.75 & 0049$-$5356 & 2019-08-16\\
4692515142570804608 & 01:22:35.90 & $-$67:59:13.95 & 17.26o0.07 & -10.09o0.05 & $0.21\pm0.03$ & 14.71 & 14.62 & 14.77 & 0122$-$6759 & 2019-08-16\\
4714790492235234176 & 02:05:56.74 & $-$59:38:35.84 & 6.14o0.07 & 2.11o0.06 & $0.15\pm0.04$ & 13.07 & 12.97 & 13.15 & 0205$-$5938 & 2019-08-20\\
5050539802435511552 & 02:45:48.67 & $-$34:18:56.36 & 48.43o0.04 & -9.93o0.06 & $0.51\pm0.04$ & 12.41 & 12.40 & 12.36 & 0245$-$3418 & 2019-08-18\\
5153423785504601216 & 02:59:01.73 & $-$17:04:59.96 & 0.67o0.06 & -6.63o0.07 & $0.17\pm0.04$ & 13.58 & 13.52 & 13.63 & 0259$-$1705 & 2019-08-20\\
5768566765625082112 & 16:21:57.93 & $-$83:50:56.96 & 2.77o0.07 & 6.87o0.08 & $0.18\pm0.04$ & 14.01 & 13.83 & 14.20 & 1621$-$8350 & 2019-08-20\\
4568776103982893056 & 17:15:28.48 & +23:41:27.05 & 1.39o0.04 & -25.74o0.05 & $0.26\pm0.03$ & 14.27 & 14.21 & 14.30 & 1715+2341 & 2019-08-18\\
6414065681237630720 & 18:39:12.47 & $-$75:15:41.08 & -0.63o0.06 & -16.08o0.09 & $0.20\pm0.05$ & 16.03 & 15.95 & 16.09 & 1839$-$7515 & 2019-08-19\\
6655084093244687872 & 18:51:05.60 & $-$50:36:39.75 & 2.62o0.06 & -10.83o0.06 & $0.16\pm0.04$ & 15.04 & 15.01 & 15.01 & 1851$-$5036 & 2019-08-19\\
6421018924052925952 & 19:00:44.83 & $-$70:32:03.13 & -12.38o0.04 & -22.21o0.05 & $0.28\pm0.03$ & 13.37 & 13.37 & 13.32 & 1900$-$7032 & 2019-08-18\\
6421018717894498048 & 19:00:33.66 & $-$70:33:35.28 & 5.70o0.05 & -2.46o0.06 & $0.15\pm0.04$ & 13.83 & 13.71 & 13.96 & 1900$-$7033 & 2019-08-20\\
6446292607565372800 & 19:19:54.92 & $-$59:38:28.88 & 5.21o0.06 & -21.40o0.05 & $0.22\pm0.05$ & 14.45 & 14.43 & 14.41 & 1919$-$5938 & 2019-08-17\\
4180161033474768640 & 19:39:22.50 & $-$16:40:43.31 & 1.43o0.11 & -8.85o0.09 & $0.23\pm0.05$ & 12.66 & 12.56 & 12.70 & 1939$-$1640 & 2019-08-19\\
6683685723576804992 & 19:44:12.51 & $-$46:28:58.22 & -0.38o0.08 & -36.87o0.07 & $0.30\pm0.05$ & 13.32 & 13.19 & 13.40 & 1944$-$4628 & 2019-08-16\\
6374275970335856512 & 20:19:00.99 & $-$71:07:53.49 & 6.13o0.05 & -37.38o0.06 & $0.39\pm0.04$ & 12.74 & 12.72 & 12.71 & 2019$-$7107 & 2019-08-17\\
6846204709077343872 & 20:24:00.16 & $-$27:25:25.22 & -11.54o0.07 & -22.39o0.05 & $0.29\pm0.04$ & 13.99 & 13.96 & 13.98 & 2024$-$2725 & 2019-08-18\\
6855907796113999872 & 20:28:33.90 & $-$22:18:19.49 & -10.38o0.10 & -28.60o0.06 & $0.31\pm0.07$ & 12.38 & 12.37 & 12.33 & 2028$-$2218 & 2019-08-17\\
6901493616919722496 & 20:34:40.08 & $-$11:08:05.94 & -13.81o0.10 & -5.86o0.06 & $0.34\pm0.06$ & 12.45 & 12.37 & 12.50 & 2034$-$1108 & 2019-08-19\\
6474092762640262016 & 20:34:20.66 & $-$53:07:06.23 & -6.54o0.06 & -18.68o0.06 & $0.19\pm0.04$ & 13.43 & 13.41 & 13.40 & 2034$-$5307 & 2019-08-17\\
6854645866004127872 & 20:40:17.89 & $-$23:36:47.22 & -23.18o0.07 & -29.89o0.05 & $0.38\pm0.04$ & 13.36 & 13.33 & 13.33 & 2040$-$2336 & 2019-08-17\\
6805531918539239040 & 20:54:36.95 & $-$25:26:12.89 & -4.22o0.09 & -5.34o0.05 & $0.18\pm0.05$ & 13.43 & 13.33 & 13.52 & 2054$-$2526 & 2019-08-19\\
6470192756241354240 & 20:58:10.82 & $-$54:07:33.00 & 1.14o0.06 & -28.15o0.06 & $0.29\pm0.05$ & 13.11 & 13.10 & 13.05 & 2058$-$5407 & 2019-08-18\\
6456376915897827456 & 21:03:25.75 & $-$58:15:35.28 & -15.18o0.06 & -20.15o0.06 & $0.26\pm0.04$ & 13.48 & 13.46 & 13.46 & 2103$-$5815 & 2019-08-18\\
6802706551612738432 & 21:10:28.12 & $-$27:00:29.36 & -2.31o0.08 & -25.71o0.05 & $0.27\pm0.06$ & 13.28 & 13.28 & 13.23 & 2110$-$2700 & 2019-08-18\\
6401710125478435200 & 21:14:02.00 & $-$64:38:47.32 & 16.33o0.04 & -17.01o0.04 & $0.24\pm0.03$ & 13.97 & 13.93 & 13.98 & 2114$-$6438 & 2019-08-17\\
6479124127849108736 & 21:19:11.03 & $-$48:27:17.63 & 16.91o0.08 & -11.78o0.06 & $0.21\pm0.05$ & 14.53 & 14.51 & 14.47 & 2119$-$4827 & 2019-08-17\\
6467153298080614400 & 21:25:20.11 & $-$48:34:50.08 & 2.13o0.06 & -19.70o0.06 & $0.19\pm0.05$ & 14.11 & 14.09 & 14.09 & 2125$-$4834 & 2019-08-20\\
6403485802397156992 & 21:27:10.30 & $-$63:00:27.09 & 12.34o0.03 & -16.52o0.04 & $0.17\pm0.03$ & 13.86 & 13.85 & 13.81 & 2127$-$6300 & 2019-08-18\\
6593248963749480320 & 21:45:10.70 & $-$30:22:26.73 & 1.00o0.07 & -25.95o0.07 & $0.25\pm0.05$ & 13.66 & 13.64 & 13.63 & 2145$-$3022 & 2019-08-20\\
6619758674426791936 & 22:07:39.31 & $-$26:45:05.56 & 11.25o0.11 & 2.16o0.09 & $0.24\pm0.07$ & 12.11 & 12.04 & 12.18 & 2207$-$2645 & 2019-08-19\\
6568272908586299136 & 22:07:29.23 & $-$43:29:30.49 & 17.90o0.04 & -25.14o0.05 & $0.21\pm0.04$ & 13.30 & 13.27 & 13.28 & 2207$-$4329 & 2019-08-19\\
6559602950163581952 & 22:07:25.53 & $-$50:21:05.38 & -8.77o0.05 & -26.38o0.07 & $0.28\pm0.04$ & 13.72 & 13.69 & 13.69 & 2207$-$5021 & 2019-08-19\\
6409382345817774080 & 22:08:51.34 & $-$60:26:11.43 & 0.49o0.04 & -12.51o0.06 & $0.27\pm0.04$ & 13.82 & 13.74 & 13.86 & 2208$-$6026 & 2019-08-19\\
6568081112526763648 & 22:13:57.49 & $-$43:57:43.72 & 8.19o0.05 & -11.90o0.06 & $0.29\pm0.04$ & 13.67 & 13.59 & 13.73 & 2213$-$4357 & 2019-08-19\\
6566677242336303360 & 22:14:14.76 & $-$46:20:06.71 & 7.46o0.05 & -16.93o0.06 & $0.17\pm0.04$ & 14.77 & 14.71 & 14.79 & 2214$-$4620 & 2019-08-16\\
6623822705625945472 & 22:39:37.70 & $-$24:55:02.94 & 55.31o0.12 & -24.94o0.12 & $0.59\pm0.08$ & 12.50 & 12.42 & 12.59 & 2239$-$2455 & 2019-08-18\\
6505572880055200896 & 22:50:15.10 & $-$56:25:17.08 & 13.58o0.04 & -16.19o0.04 & $0.15\pm0.03$ & 14.42 & 14.41 & 14.38 & 2250$-$5625 & 2019-08-20\\
6514882651165295232 & 22:59:35.83 & $-$49:31:18.23 & 8.68o0.04 & -27.30o0.06 & $0.27\pm0.06$ & 14.29 & 14.28 & 14.27 & 2259$-$4931 & 2019-08-20\\
6543384947494107392 & 23:04:43.81 & $-$40:21:07.52 & 2.04o0.05 & -5.30o0.06 & $0.14\pm0.04$ & 13.89 & 13.86 & 13.87 & 2304$-$4021 & 2019-08-20\\
6556633886515772544 & 23:07:12.20 & $-$31:57:38.11 & 20.57o0.10 & -9.06o0.07 & $0.35\pm0.05$ & 12.30 & 12.24 & 12.36 & 2307$-$3157 & 2019-08-19\\
6493693515910526208 & 23:09:26.75 & $-$56:23:02.24 & 8.79o0.05 & -9.18o0.06 & $0.32\pm0.04$ & 13.53 & 13.47 & 13.56 & 2309$-$5623 & 2019-08-20\\
6391197660443817984 & 23:09:32.88 & $-$67:29:14.43 & 6.16o0.04 & -23.25o0.05 & $0.20\pm0.03$ & 13.72 & 13.70 & 13.70 & 2309$-$6729 & 2019-08-18\\
6503103617457504000 & 23:13:52.61 & $-$48:58:15.92 & 15.52o0.03 & -10.51o0.04 & $0.19\pm0.03$ & 14.09 & 14.08 & 14.04 & 2313$-$4858 & 2019-08-18\\
6389351133744091648 & 23:48:08.84 & $-$66:12:17.59 & 2.88o0.05 & -9.08o0.05 & $0.15\pm0.03$ & 15.41 & 15.38 & 15.40 & 2348$-$6612 & 2019-08-20\\
\hline
\end{tabular}
\tablefoot{The short name is based on the {\em Gaia} RA and Dec, expressed in sexagesimal units. 1900$-$7033 and 2239$-$2455 were previously known as \object{JL\,6} and \object{PHL\,5382}, respectively.}
\label{t:phot}
\end{table*}

\section{Data analysis}
\label{sec:analysis}
For the atmospheric characterization of the observed targets, we follow a three-step approach. First, we perform model fitting of their entire spectral energy distributions (SED) that consist  of the available ground-based and space-borne  broad-band photometry, from which we estimate the  atmospheric parameters of each star, i.e., effective temperature and surface gravity, as well as the interstellar reddening and angular diameter ($T_{\rm eff}$, $\log{g}$, $E(44-55)$, and $\Theta$). Here, $E(44-55)$ is the monochromatic reddening measured at 4400 and 5500\,\AA, which is closely related to the more traditional band-integrated color excess $E(B-V)$, following the definition of \citet{fitzpatrick2019}. Second, we use the photometric estimates as initial guesses for the spectral analysis of flux calibrated spectra. Third, we re-determine $E(44-55)$ and $\Theta$, by employing the spectroscopic best-fit model in the SED-fitting procedure.

\subsection{Grids of synthetic spectra}
\label{sec:grids}

Due to the low resolution of the spectra in hand, only stellar parameters such as $T_{\rm eff}$ and $\log{g}$ can reliably be measured, while rotation, metallicity, and helium abundance are very poorly constrained, especially in the late-B/early-A spectral type transition, where many candidates are found. In this region of the Hertzsprung-Russell diagram, MS and BHB stars can not be distinguished on the basis of their atmospheric parameters alone (see Fig. \ref{fig:hr}), but additional information on the chemical composition and rotation of the stars is needed \citep[see e.g.][]{2008ASPC..392..167H}. Because the lack of spectral resolving power does not allow us to accurately obtain this auxiliary information, we perform the analysis of the SEDs and spectra for all targets twice by using two different grids of synthetic models, one being appropriate for MS stars and one being tailored to BHB stars, see below. The distinction between MS and BHB nature is then made based on the agreement of their spectrophotometric and parallax-based distances (see Sect.~\ref{sec:parameters}).

For MS stars it is self-evident to adopt a solar chemical composition. In contrast, the abundance patterns of BHB stars are diverse. Because the targets were selected to have high tangential velocities, BHB candidates are likely halo stars and thus metal poor. However, the stars' chemical composition also varies along the horizontal branch. The morphology of BHBs of globular clusters reveals a universal discontinuity at $T_{\rm eff} \sim 11\,500$\,K \citep{1999ApJ...524..242G,2016ApJ...822...44B}, which coincides with a drastic change in the chemical composition.
While BHB stars cooler than  11\,500K have the same metal content as the bulk of the stars of the host cluster, the abundance pattern of the hotter stars is peculiar, with helium depletion and iron enrichment \citep{behr2003}. These chemical pecularities have been attributed to atomic diffusion, that is, to an interplay of gravitational settling and radiative levitation. The former leads to helium depletion and the latter causes the iron enrichment. The onset of atomic diffusion coincided with the disappearance of surface convection \citep{2002HiA....12..292S,2011A&A...529A..60M}.
Because we can not determine the abundance pattern of the individual BHB stars, we calculate model atmospheres and synthetic spectra assuming a metalicity of one tenth solar for all BHB stars, but different helium abundances, that is, solar helium content
for stars cooler than 11\,500\,K and 1/10 solar for the hotter ones.

Both grids span $T_{\rm eff} = 7200$--$33\,000$\,K and $\log{g} = 3.0$--$4.6$.
For $T_{\rm eff} \geq 9000$\,K, we adopt the hybrid approach ``ADS'' \citep{przybilla2011, irrgang2014}, which combines {\sc Atlas12} \citep{kurucz1996} to compute hydrostatic model atmospheres in local thermodynamic equilibrium (LTE) with {\sc  Detail} \citep{giddings1981} to account for non-LTE effects; the {\sc Surface} software \citep{giddings1981} is used to produce the synthetic spectrum. Another set of models, which extends below $10\,000$\,K, has been generated by accounting only for LTE by means of the {\sc Atlas12} and {\sc synthe} software \citep{kurucz1993}. All models include the recent improvements introduced by \citet{irrgang2018a} for the computation of level dissolution in proximity of the Balmer jump. 
The synthetic photometry is based on SEDs computed with {\sc Atlas12} using a ten times finer frequency grid, which were convolved with the appropriate transmission profile of each 
band-pass and subsequently shifted to the respective zero point.

   \begin{figure*}
   \centering
  \includegraphics[width=0.3\linewidth]{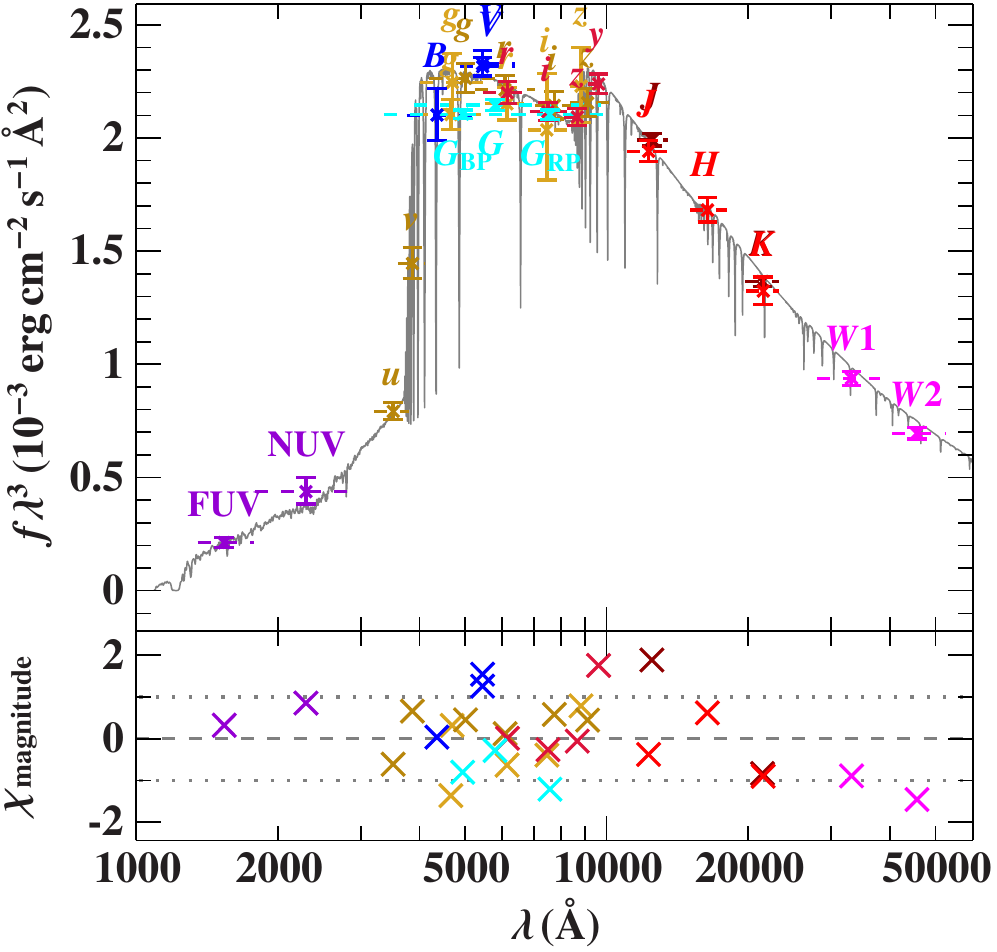} 
   \includegraphics[width=0.3\linewidth]{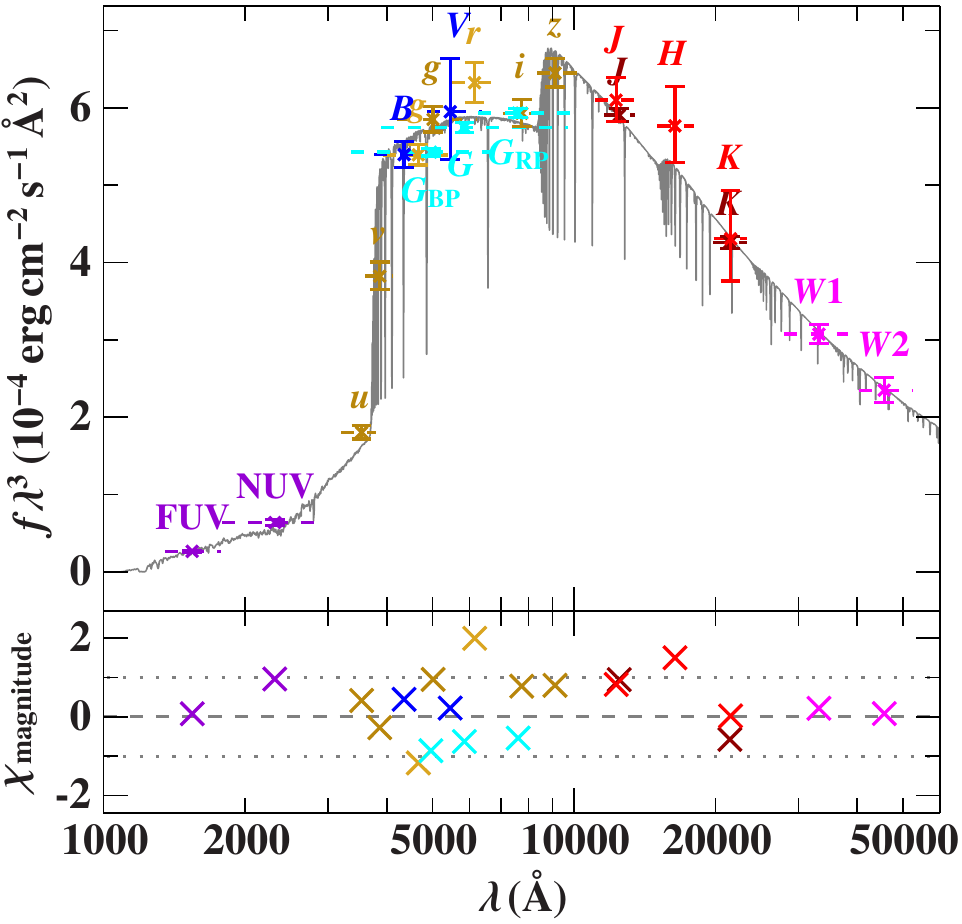}  
  \includegraphics[width=0.3\linewidth]{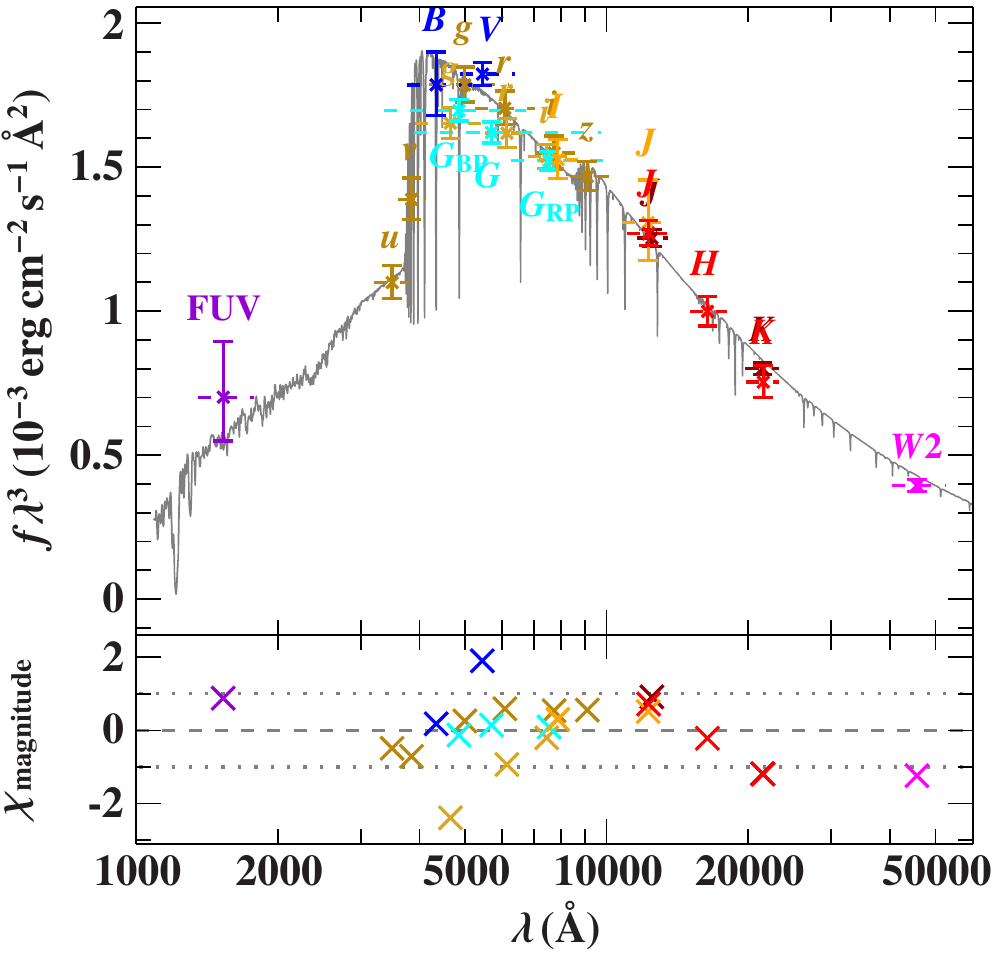} 
 \includegraphics[width=0.3\linewidth]{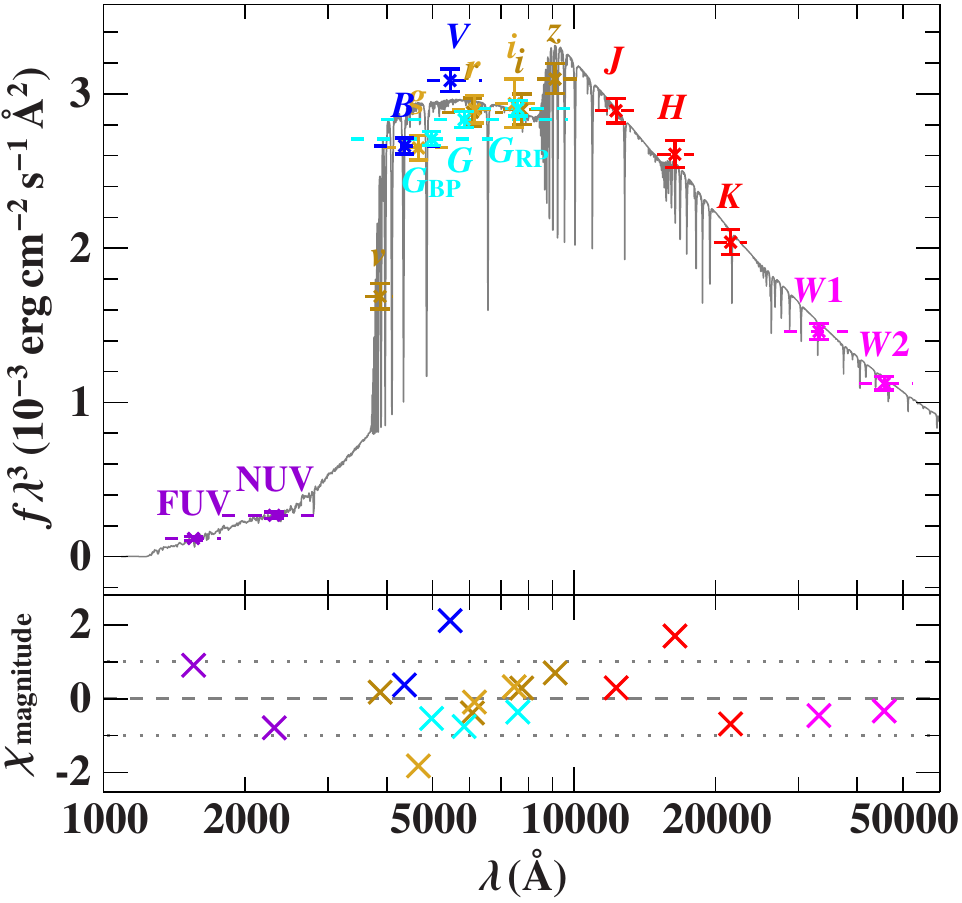} 
\includegraphics[width=0.3\linewidth]{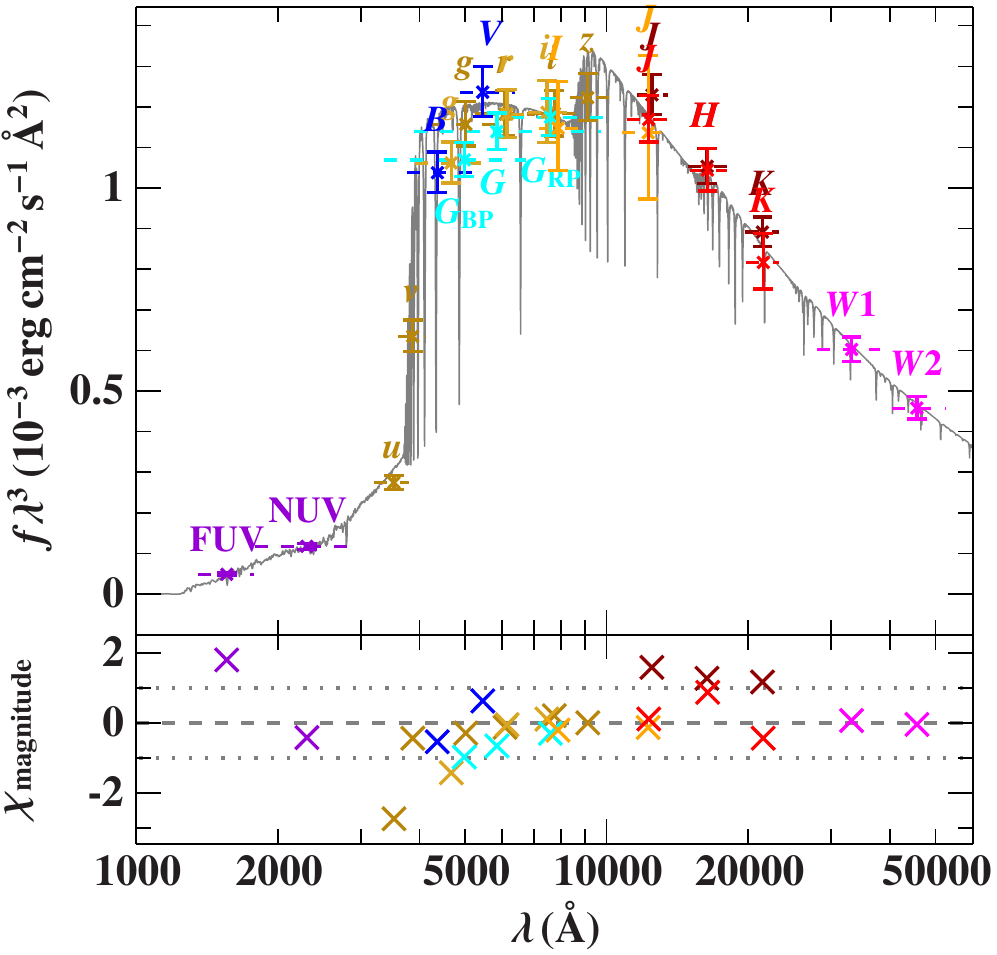} 
\includegraphics[width=0.3\linewidth]{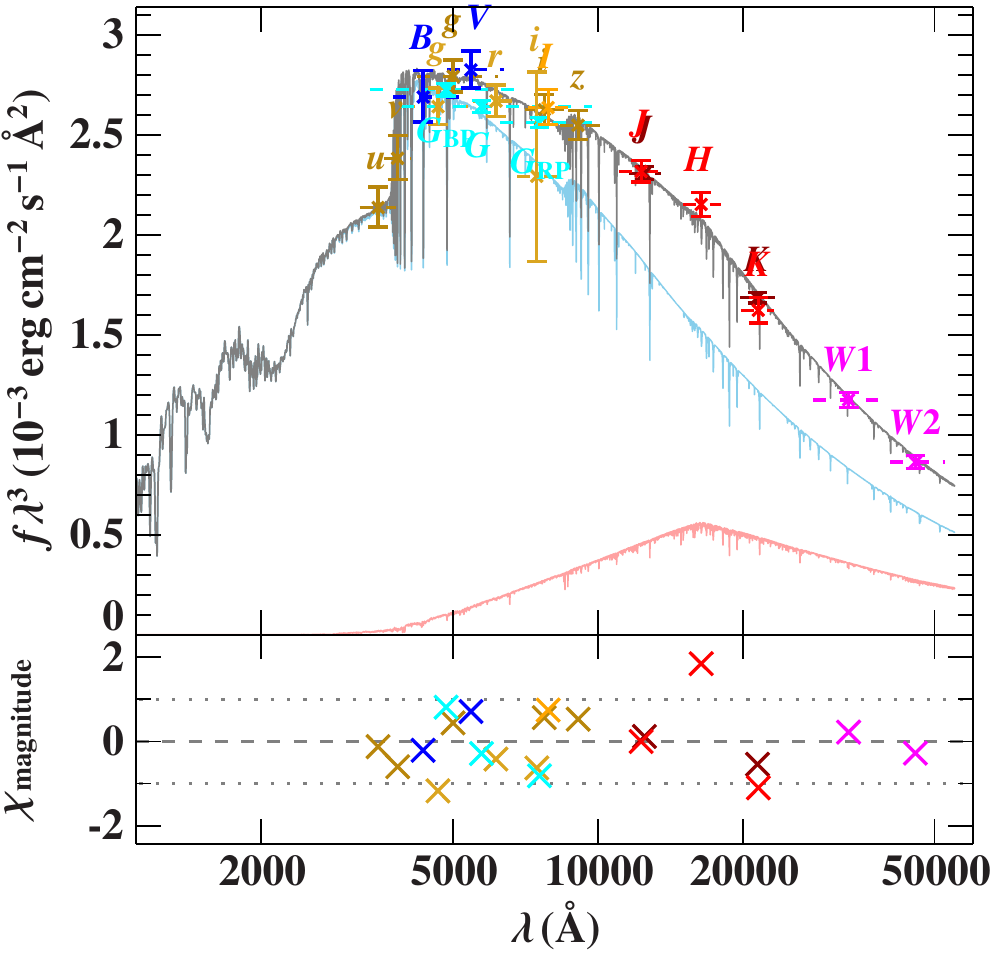}

        \caption{Examples for SED fits, from top left to bottom right: the MS candidates 0259$-$1705, 1900$-$7033, 2207$-$4329, and 2259$-$4931; the BHB candidate 1851$-$5036; and the binary 1944$-$4628. In each plot, the top panels show the measurements (color-coded by photometric system) as well as the best-fitting model spectra (gray), while the bottom panels show the error-normalized best-fit residuals, $\chi = (O_i - M_i)/\sigma_i$, for each band-pass. The object 1944$-$4628 requires a composite SED fit, which is the combination of a $T_{\rm eff} \approx 24\,400$\,K and a $T_{\rm eff} \approx 5900$\,K component (blue and red curve, respectively).}
    \label{fig:phot-fit}
    \end{figure*}
\subsection{SED fitting}

We use the {\sc stilts} software \citep{taylor2006} to query all the major photometric surveys that are available through the Table Access Protocol (TAP) of the VizieR online service \citep{vizier}, which is hosted by the Strasbourg Astronomical Data Center. For most of the stars, in addition to {\em Gaia} photometry, there is reliable coverage from the ultraviolet \citep[{\em GALEX};][]{bianchi2017} through the optical \citep[APASS and SkyMapper;][as well as SDSS and PanSTARRS, for targets  with $\delta \gtrsim -30$\,deg; \citealt{alam2015}; \citealt{chambers2016}]{henden2015,wolf2018} and infrared \citep[VHS, 2MASS, and {\em WISE};][]{mcmahon2013,skrutskie2006,2014yCat.2328....0C,2019ApJS..240...30S}
We adopt the most recent calibration of {\em GALEX} \citep{wall2019}, well-validated zero points and uncertainties for the other photometric surveys\footnote{SkyMapper \citep{casagrande2019}, SDSS \citep{holberg2006}, PanSTARRS \citep{tonry2012}, 2MASS \citep{maiz-appelaniz2018b}, UKIDDS \citep{hewett06}, VHS \citep{gonzalez-fernandez2018}, WISE \citep{jarrett2011}.}, as well as suggested corrections of {\em Gaia} DR2 photometry and the corresponding filter transmission profiles \citep{evans2018,maiz-apellaniz2018}. 

The SED fitting  procedure follows  the methods  outlined by \citet{heber2018}, allowing to directly estimate the four parameters: $\Theta$, $T_{\rm eff}$, $\log{g}$, and $E(44-55)$. We adopt the monochromatic extinction law by \citet{fitzpatrick2019} with the total-to-selective extinction coefficient $R(55) = 3.02$, as typical for the interstellar medium. The best-fit is found by matching observed and synthetic photometry via $\chi^2$ minimization. Each stellar SED is fitted with both model grids described in the previous section. Six examples are displayed in Fig.~\ref{fig:phot-fit}.

The preliminary SED fitting does not only provide an initial guess for the spectroscopic analysis, but is also crucial for identifying problematic objects like binary stars that exhibit composite SEDs. Indeed, we find that a single stellar component is not sufficient to model the SED of 1944$-$4628, which is likely an evolved low-mass star with an infrared excess due to a cooler binary companion (Fig.~\ref{fig:phot-fit}). Last but not least: photometric temperature estimates provide an important consistency check for our spectroscopic analysis, see the following section.

\begin{figure*}
    \centering
    \includegraphics[width=\linewidth]{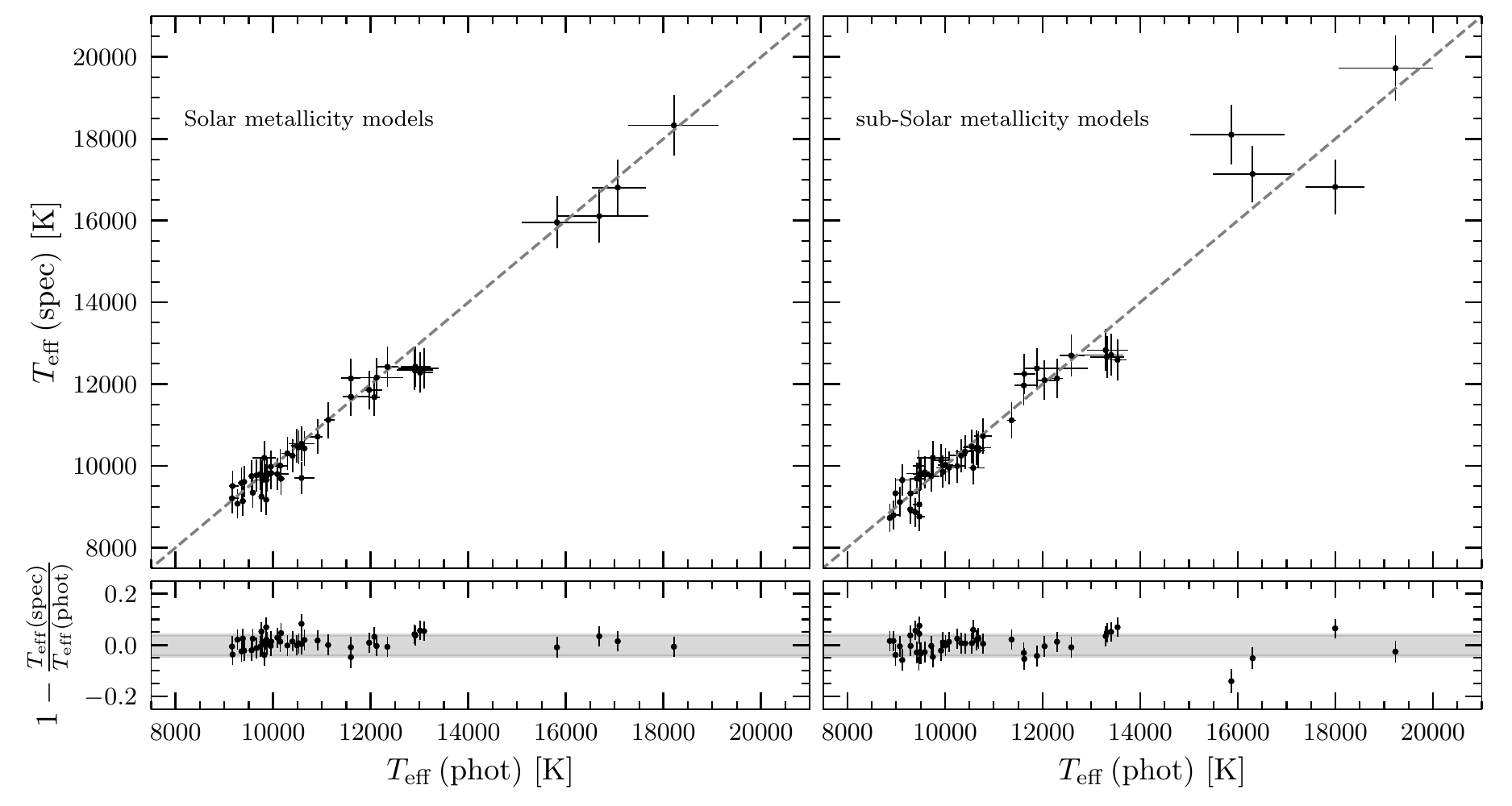}
    \caption{Comparisons between photometric $T_{\rm eff}$, measured via SED fitting, and spectroscopic $T_{\rm eff}$. The result for MS (top left panel) and BHB models (top right panel) are shown. The dashed lines represent the equality curves. The residuals are displayed as function of photometric $T_{\rm eff}$ in the lower panels. The gray shaded area represent the 4\% systematic uncertainty assumed for the spectroscopic results.}
    \label{fig:teff_comparison}
\end{figure*}
   \begin{figure*}
   \centering
   \includegraphics[width=17cm]{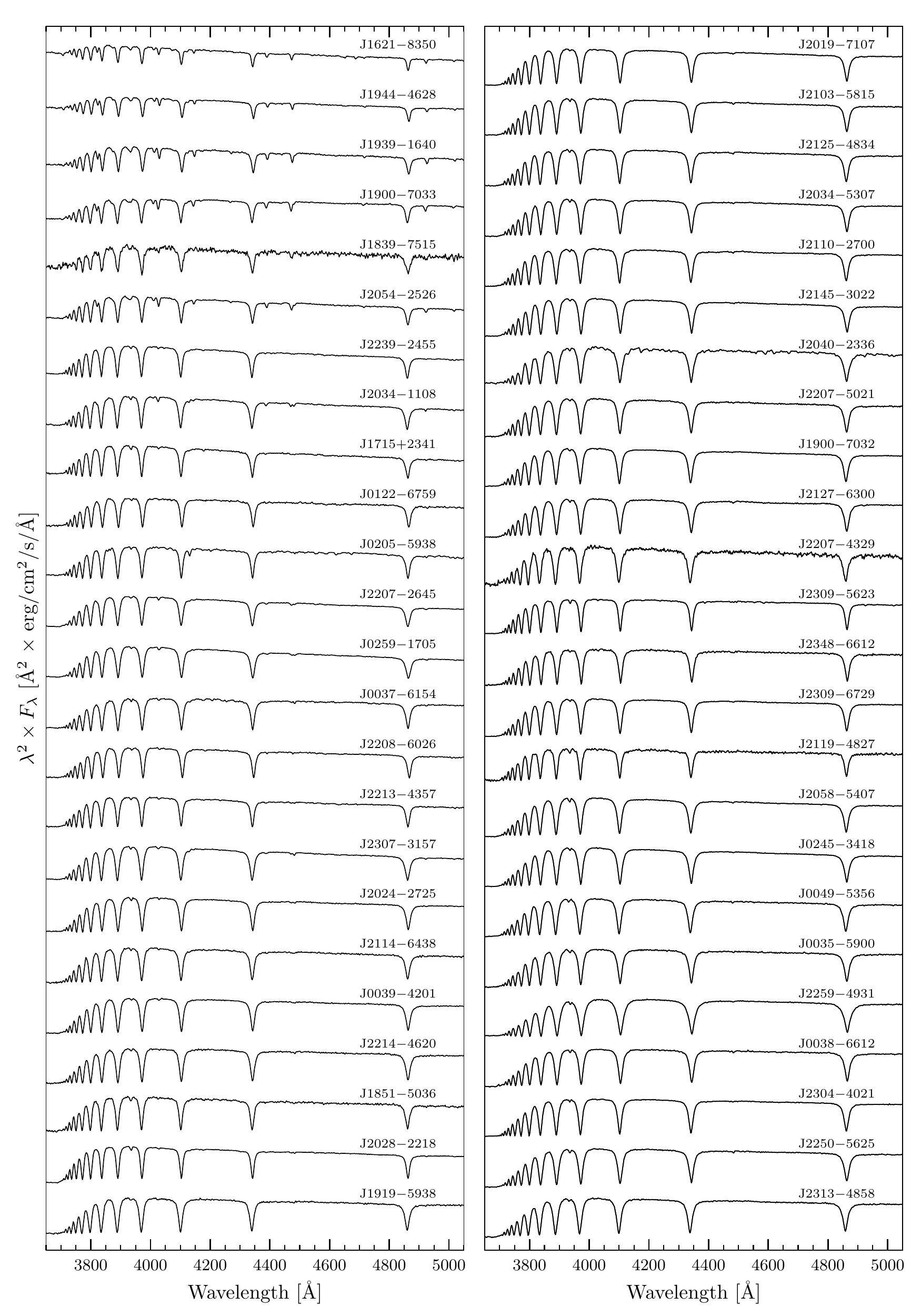}
   \caption{Observed spectra, sorted by decreasing $T_{\rm eff}$.}
    \label{fig:spectra}
    \end{figure*}
   \begin{figure*}
   \centering
   \includegraphics[width=0.49\linewidth]{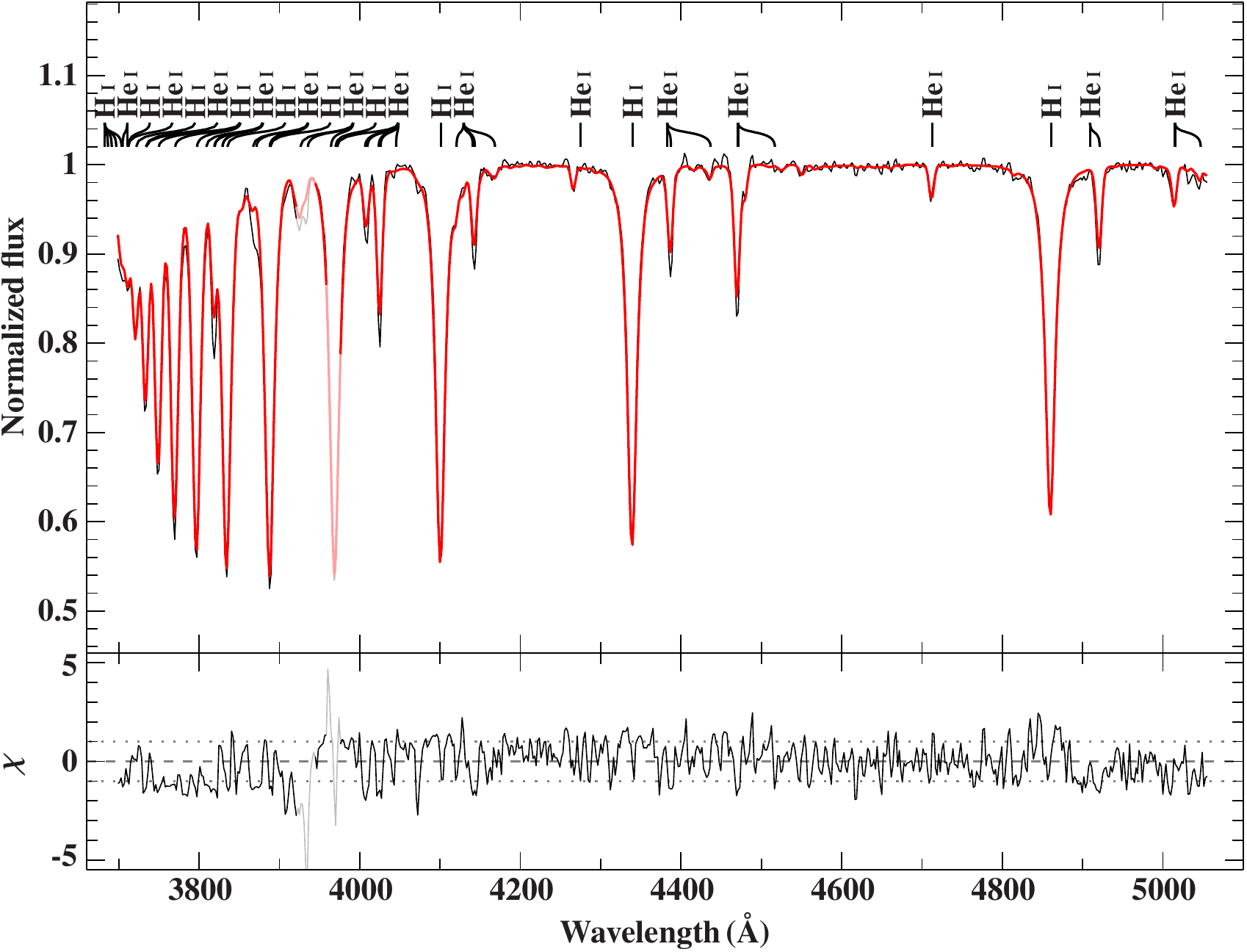}
   \includegraphics[width=0.49\linewidth]{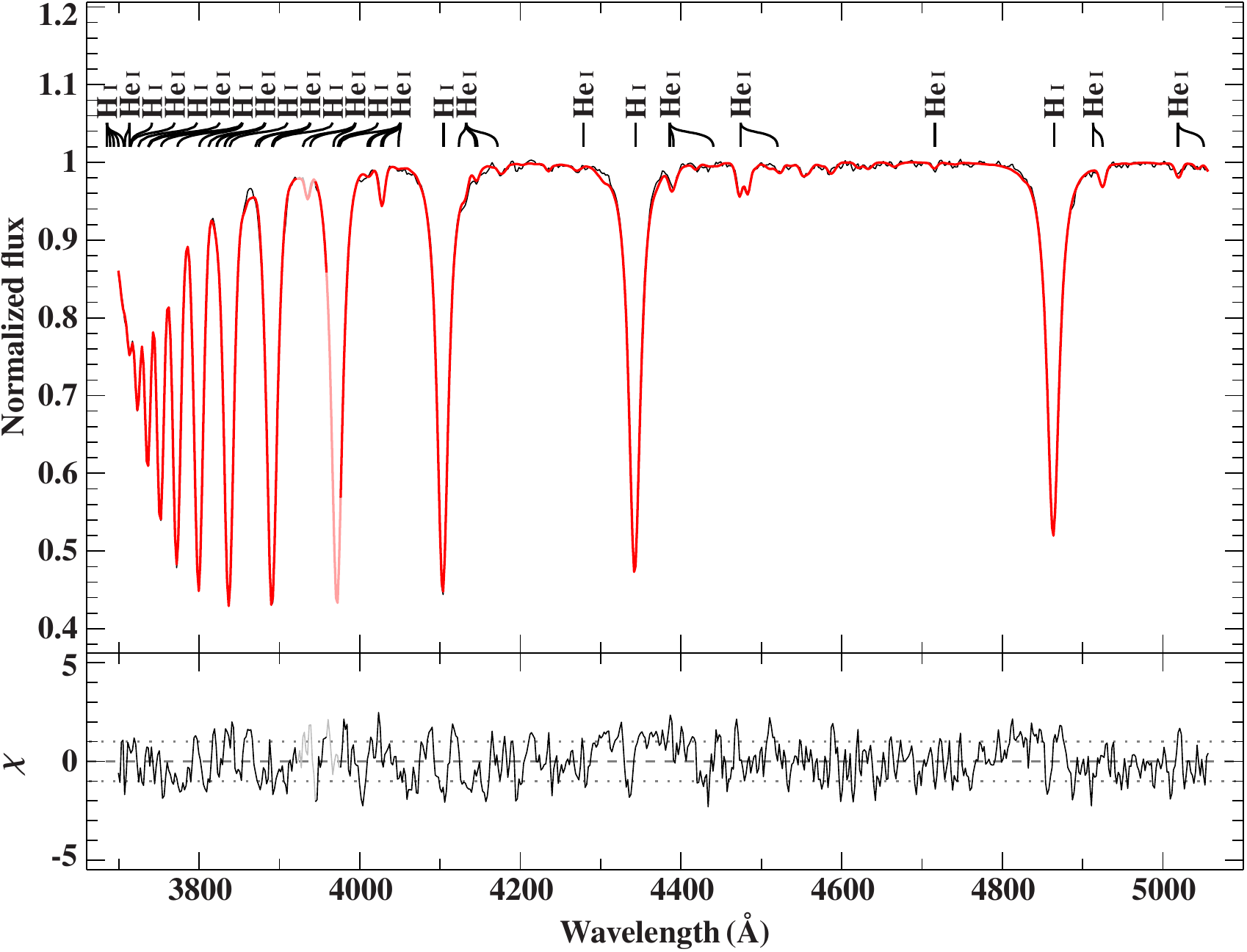}  
   \includegraphics[width=0.49\linewidth]{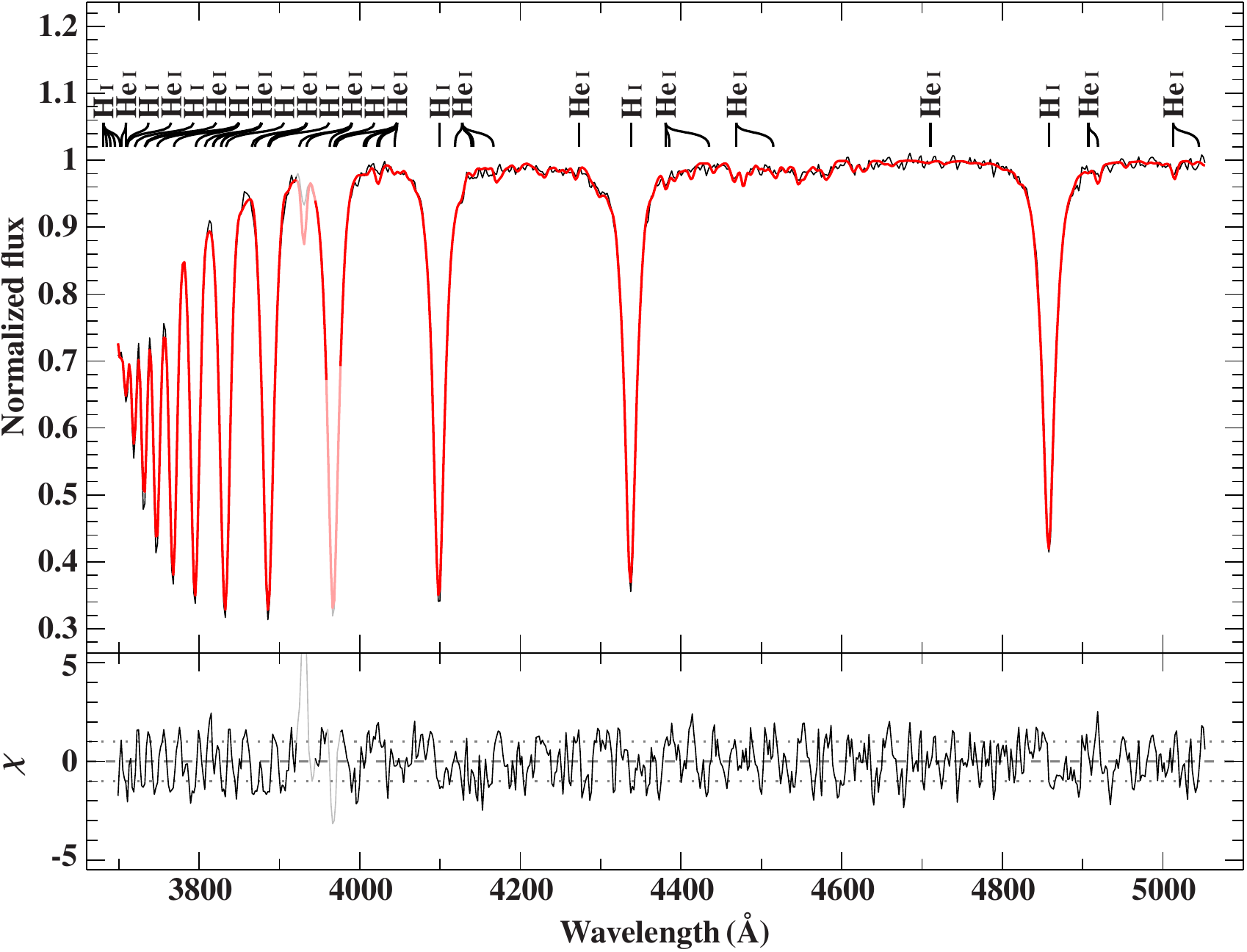}   
   \includegraphics[width=0.49\linewidth]{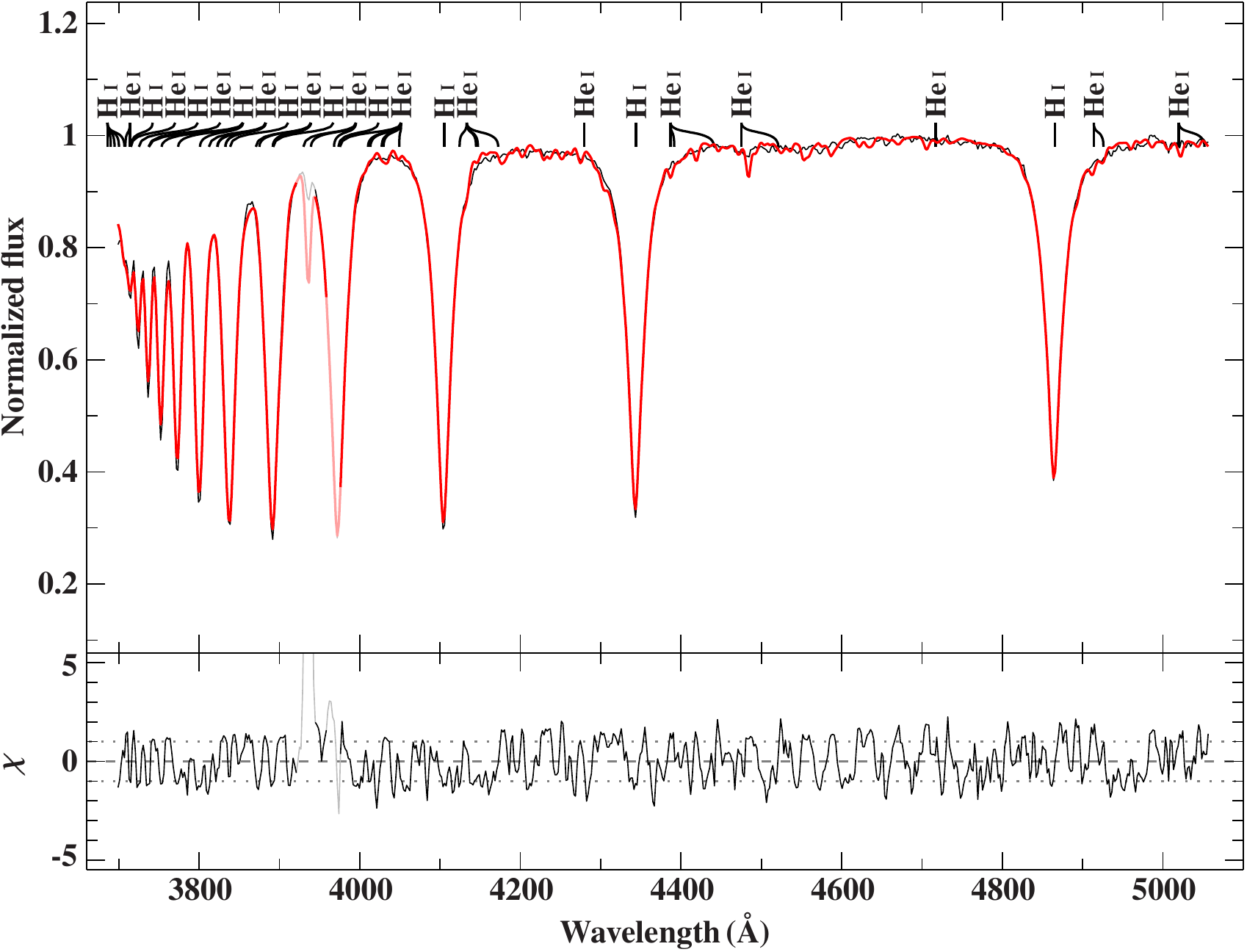}
   \caption{Examples of spectroscopic best-fits for four MS candidates (Sect.~\ref{sec:parameters}) with error-normalized residuals; from top left to bottom right: 1900$-$7033, 0259$-$1705, 2207$-$4329, and 2259$-$4931. The normalized spectra and the best-fit solar-metallicity models are shown in black and red, respectively. The two top panels show ADS models, while the bottom ones are obtained from the {\sc Atlas12/synthe} grid. H and He lines are labeled at the corresponding wavelengths. The Ca\,{\sc ii} lines are masked out during the fitting procedure due to possible interstellar contamination, which is recognizable in the spectrum of 1900$-$7033; masked regions are plotted using lighter colors.  We note that the He\,{\sc i} lines of 1900$-$7033 are stronger than those of our solar-metallicity model (c.f. Sect.~\ref{sec:spec-fitting} for a discussion on this star). The second star, 0259$-$1705, shows an excellent agreement between observed and model spectrum. In contrast,  the Ca\,{\sc ii}\,H lines of the two coolest stars in the bottom panels and  the Mg\,{\sc ii} 4481 line of 2259$-$4931 are weaker than those of the best-fit MS model. This issue, although to a lesser extent, also characterizes the best-fit obtained with sub-solar metallicity models.}
    \label{fig:best-fit}
    \end{figure*}
\subsection{Spectroscopic fitting}
\label{sec:spec-fitting}
The analysis of flux-calibrated spectra follows the methods outlined by \citet{irrgang2014}, where the best-fits are determined via $\chi^2$ statistics through comparison with our model grids. We preferentially use ADS models, but we switch to the {\sc Atlas12/synthe} grid in order to avoid extrapolation at the edge of the ADS grids. We ignore macroturbulence and assume a microturbulence of 2\,km\,s$^{-1}$. As stated in Sect.~\ref{sec:grids},  the low resolution of our spectra makes it difficult to measure detailed metal abundances and stellar rotation, hence, our approach is to fit the spectra with both the solar composition and the metal-deficient grids, respectively.

Mg and Si are the only elements for which the individual abundances are allowed to independently vary when using the ADS grids. However, we note that most measurements are poorly constrained and do not deliver reliable abundances, which is why we refrain from giving those numbers. Although we keep the projected rotational velocity, $\varv \sin{i}$, as a free parameter, it can be hardly resolved from our low-resolution spectra ($\Delta \varv \approx 400$\,km\,s$^{-1}$). The spectra are corrected for the barycentric velocity and the radial velocity, $\varv_{\rm rad}$, is independently determined from each exposure. Statistical uncertainties affecting $T_{\rm eff}$ and  $\log{g}$ are numerically inconspicuous. The systematic uncertainties dominating the error budget of $T_{\rm eff}$ and  $\log{g}$ are estimated to the level of 4\%  and 0.1\,dex, respectively. The error budget of $\varv_{\rm rad}$ is also dominated by systematic uncertainties, which are estimated from the difference between the measured values of the two available exposures for each star ($\lesssim 10$\,km\,s$^{-1}$) and the 25\,km\,s$^{-1}$ error, determined by comparing the $\varv_{\rm rad}$ of the white dwarf standards to values from the literature (Sect.~\ref{sec:observations}).

The main results of our analysis are presented in Fig.~\ref{fig:teff_comparison}, which displays the comparison between photometric and spectroscopic $T_{\rm eff}$ for the MS and BHB model analyses. The good agreement seen in Fig.~\ref{fig:teff_comparison} confirms the robustness of the SED and spectroscopic analyses, but also suggests that an additional method (like the distance comparison discussed in the following section) is necessary to distinguish between MS and BHB candidates. 

Two stars, 1621$-$8350 and 1944$-$4628, are considerably hotter than the bulk of the sample, and are not included in Fig.~\ref{fig:teff_comparison}. 1621$-$8350 is the only star that shows a He\,{\sc ii} line (at 4685.8\,\AA) in its spectrum (Fig.~\ref{fig:spectra}), which leads to the classification of an O type star that is consistent with a high $T_{\rm eff} \approx 30\,000$\,K. 1944$-$4628 is an early B-type star in a binary system (Fig.~\ref{fig:phot-fit}) at $T_{\rm eff} \approx 24\,400$\,K. Both stars are discussed separately in Sect.~\ref{sec:parameters}.  

A spectral atlas of our program stars is presented in Fig.~\ref{fig:spectra}, while examples of spectroscopic best-fits are shown in Fig.~\ref{fig:best-fit} for four stars (0259$-$1705, 1900$-$7033, 2207$-$4329, and 2259$-$4931) that are later classified as MS candidates based on their spectrophotometric distances (Sect.~\ref{sec:parameters}). In this figure, we exhibit three typical temperature regimes of our sample. We note that for hot stars like 1900$-$7033, shown in Fig.~\ref{fig:best-fit}, and 1939$-$1640 and 2054$-$2526, it would be in fact possible to have more reliable He abundance measurements rather than assuming solar composition, but this exercises is beyond the scope of the paper. For cool objects like 2207$-$4329 and 2259$-$4931, we can see just a few weak metal lines. In these specific cases, both MS and BHB models seem to overestimate the intensity of the Ca\,\textsc{ii}\,H line, hinting at a sub-solar abundance that could be better investigated with higher-resolution spectra. The intensity of the Mg\,\textsc{ii} 4481 line of 2207$-$4329 is also overestimated. The star with an intermediate temperature, 0259$-$1705, shows an excellent agreement with the model.

Finally, we report that three stars were previously studied by other authors:\\ 

{\em 1900$-$7033:} \citet{kilkenny1989} determined $T_{\rm eff} = 18\,000$\,K  and $\log{g} = 4.0$ with 10\,\% errors from 3\,\AA-resolution spectra, in  agreement with our results. 
This author proposed the star to be mildly He-rich  ($n$He/$n$H\,$\sim 0.3$). This finding is supported by our observations, because the He lines appear slightly deeper than those of the best-fit model with solar $\log{({\rm He/H})} = -1.01$ (see Fig.~\ref{fig:best-fit}). \citet{kilkenny1989} also measured $\varv\sin{i} = 175$\,km\,s$^{-1}$ from the He lines, while the H$\gamma$ and H$\delta$ fits delivered smaller $\varv\sin{i} \approx 75$ and $100$\,km\,s$^{-1}$, respectively. If we keep He as a free parameter, the best-fit result favors slightly super-solar He, but the line cores of the observed spectrum are still deeper than the model. Given that we cannot disentangle the effect of stellar rotation due to the low resolution of our spectra, we adopt the atmospheric parameters obtained for solar He abundance. Finally, \citet{kilkenny1989} reported  a previous measurement of $\varv_{\rm rad} \approx -95$\,km\,s$^{-1}$ that is at 2\,$\sigma$ from our result.

{\em 2239$-$2455:} \citet{kilkenny1987} used Str\"omgren photometry to classify this star as B7V, which is compatible with our spectral analysis. \citet{ortiz2007} obtained 1.2\,\AA-resolution spectra over the H$\beta$--H$\epsilon$ spectral range, from which they suggest  $T_{\rm eff} > 17\,000$\,K  and $\log{g} = 4.5 \pm 0.5$ via measurements of equivalent widths and depths of the Balmer lines. 

{2307$-$3157:} \citet{ortiz2007} determined $T_{\rm eff} = 8500$--10\,000\,K and $\log{g} = 3$--3.5 with errors  of 500\,K and  0.5\,dex, respectively, that roughly agree with our results. 

   \begin{figure}
   \centering
   \includegraphics{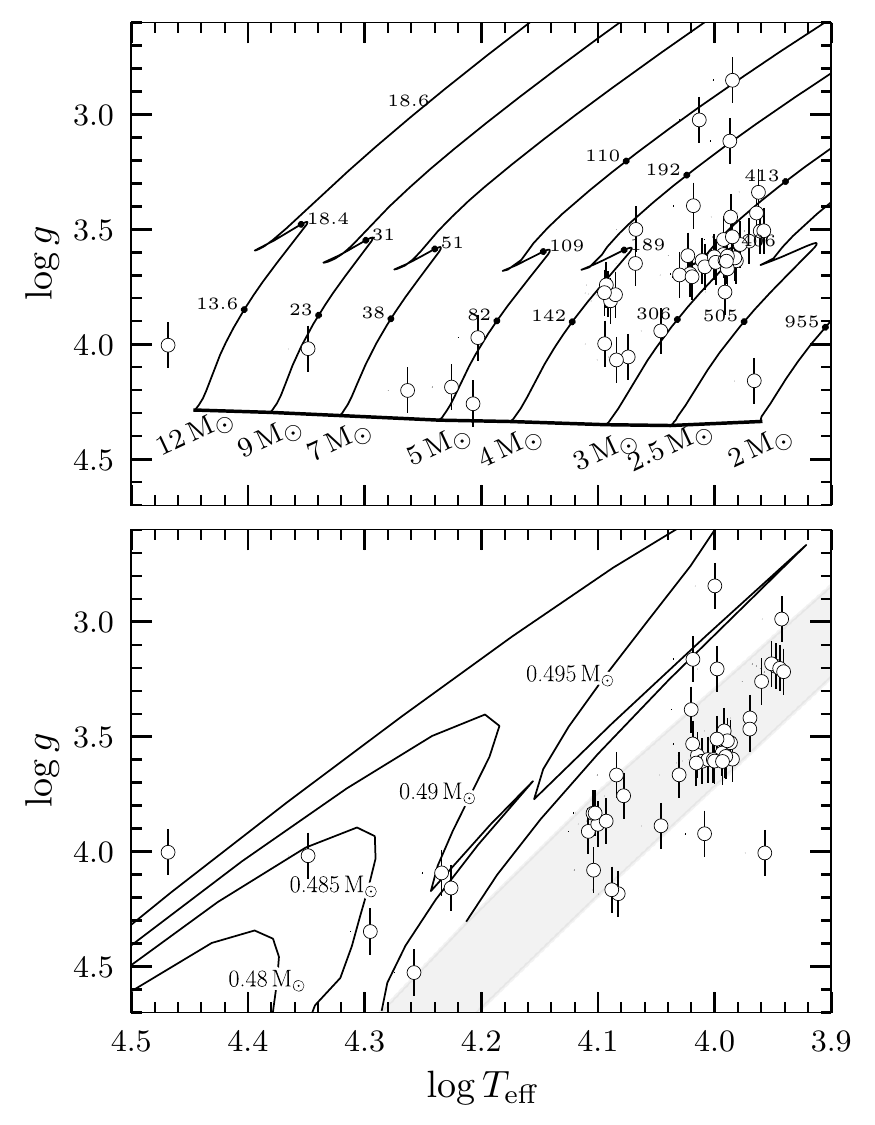}
   \caption{Kiel diagrams showing the atmospheric parameters of the observed stars as determined via the spectroscopic best-fit with model spectra representing MS and BHB stars (top and bottom panels, respectively). {\em Top panel}: Evolutionary tracks for rotating MS stars are plotted in black \citep{ekstrom2012}. The initial masses of each model are noted below the zero-age MS (thick solid curve), while the stellar ages for selected evolutionary points along the tracks are given in mega years. {\em Bottom panel}: The horizontal branch is represented by the gray shaded area, and the evolutionary tracks  for post-horizontal branch stars of different masses are drawn as solid curves \citep{dorman1992}.}
    \label{fig:kiel}
    \end{figure}
   \begin{figure*}
   \centering
   \includegraphics[width=0.9\textwidth]{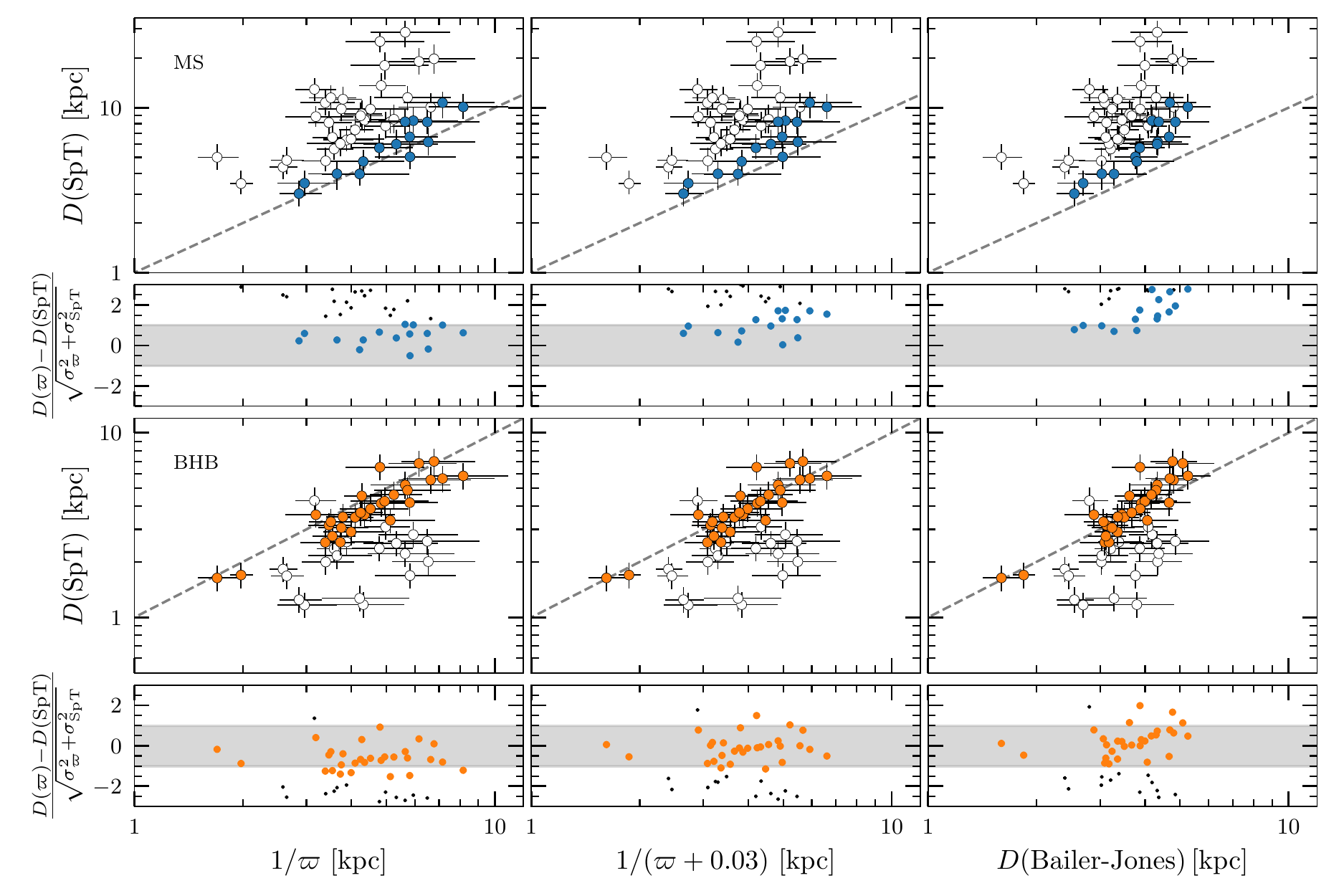}
   \caption{Comparison between spectrophotometric distances and three different  parallax-based distance estimates. {\em Upper panels}: MS candidates (blue colored symbols) are identified via the 1\,$\sigma$ agreement, which is represented by the dashed equality line and the shaded gray area. {\em Lower panels}: as for the upper panels, BHB candidates are represented by orange filled symbols. Note that three stars are classified both as MS and BHB candidates, hence they are color-colored in both the upper and lower panels.}
    \label{fig:distance}
    \end{figure*}

\subsection{Distance-driven classification and physical parameters}
\label{sec:parameters}
   \begin{figure}
   \centering
   \includegraphics{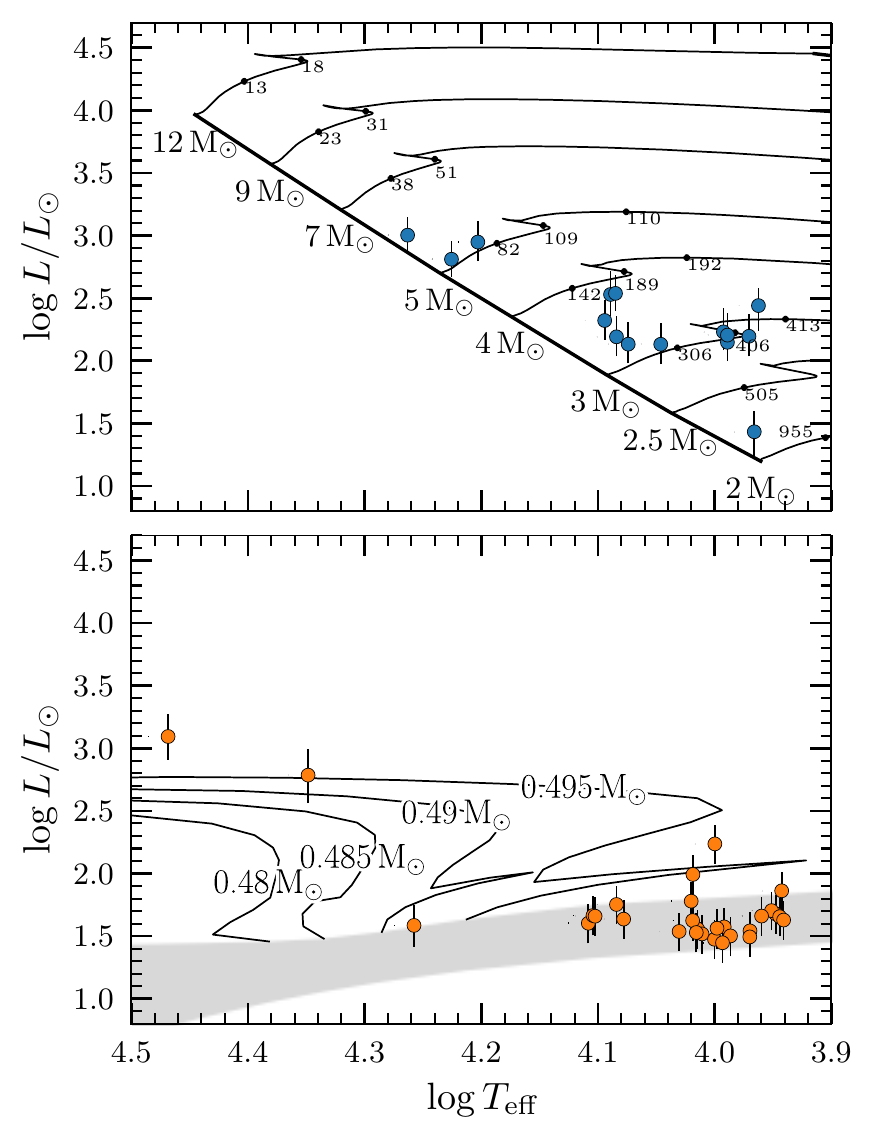}
   \caption{Theoretical Hertzsprung-Russell diagrams. The top panel shows our candidate MS stars over-plotted on \citet{ekstrom2012} evolutionary tracks for rotating stars. The masses of each model are plotted in correspondence of the zero-age MS. The luminosities are derived by adopting the spectrophotometric distances. The bottom panel shows our candidate BHB stars over-plotted on the horizontal branch that is represented by a grey-shaded area and the evolutionary tracks for post-horizontal branch stars of given masses \citep{dorman1992, dorman1993}. Note that three stars are classified both as MS and BHB candidates, hence they feature in both panels with the appropriate parameters.}
    \label{fig:hr-theo}
    \end{figure}
Figure~\ref{fig:kiel} displays the $T_{\rm eff}$ and $\log{g}$ distribution of the observed stars in the Kiel diagram,as derived from the spectral analysis with synthetic spectra representing MS and BHB stars. For reference, we plot the theoretical evolutionary tracks for MS rotating stars \citep{ekstrom2012}, and the horizontal branch and post-horizontal branch tracks \citep{dorman1992,dorman1993}. Given the degeneracy between MS stars of 3--4\,M$_{\sun}$ and BHB stars, whose evolutionary tracks overlap in the Kiel diagram, and the lack of additional information such as  $\varv\sin{i}$ and metal abundances, it is a priori impossible to classify the majority of stars in our sample by just using $T_{\rm eff}$ and $\log{g}$. 

Hence, to assess the nature of our program stars, we proceed in a different fashion, by comparing the {\em Gaia} parallaxes to the spectrophotometric distances, $D({\rm SpT})$, determined under the assumption that the observed stars are either MS or BHB stars. In order to estimate $D({\rm SpT})$ under the MS-star assumption, we first determine the stellar masses, radii, and luminosities via interpolation of the \citet{ekstrom2012} tracks shown in Fig.~\ref{fig:kiel} onto the $T_{\rm eff}$ and $\log{g}$ obtained via the spectroscopic fitting with the solar-metallicity grid.  The spectrophotometric distance is then computed from the resulting stellar radius and the angular diameter determined via SED-fitting: $D({\rm SpT}) = 2R/\Theta$. In the BHB-star assumption, we assume a canonical BHB-star mass of $M = 0.5 \pm 0.1$\,M$_{\sun}$ from which we infer the stellar radius as $R = \sqrt{G M/g}$, where $G$ is the gravitational constant and $g$ is the surface gravity measured with the representative metal-poor model grid. Then, the $D({\rm SpT})$ is calculated as above using the appropriate $\Theta$.

Figure~\ref{fig:distance}  displays the  comparison between parallax-based distance estimators, $D(\varpi)$, and spectrophotometric distances for the MS and BHB results. The simplest estimator that we consider is $D(\varpi) = 1/\varpi$, where  $D(\varpi)$ is the median of the $1/\varpi$ distribution. We also consider the effect of a global zero-point shift of +0.03\,mas, identified by \citet{lindegren2018} from the analysis of an all-sky sample of extra-galactic sources -- but these authors also noted that such an offset has relatively large variations over degree-wide scales. Finally, as advised by \citet{bailer-jones} and \citet{luri2018}, we also compare their prior-based distance estimate, $D$(Bailer-Jones), against $D({\rm SpT})$. 

MS and BHB stars are expected to lie on the identity lines depicted in Fig.~\ref{fig:distance}, when the $D({\rm SpT})$ computed from the corresponding models agree with the parallax-based distance estimates. From this comparison, we identify 15 MS candidates that have $D({\rm SpT})$ in agreement within 1\,$\sigma$ with one of the three $D(\varpi)$ estimators (top panels of Fig.~\ref{fig:distance}). The same comparisons for BHB estimates deliver 29 candidates (lower panels in Fig.~\ref{fig:distance}). We note that three stars (0035$-$5900, 2127$-$6300, and 2304$-$4021) are both identified as MS and BHB candidates, while all the others are uniquely classified either as MS or BHB candidates. Among the 29 BHB candidates, there are 1621$-$8350 and 1944$-$4628 too, which we discuss in more detail below as evolved extreme horizontal branch stars (post-EHB). Seven stars of our sample do not fit in any of the two proposed classification schemes. 

In general, we note that all MS candidates agree with the $1/\varpi$ estimator and that the other two estimators tend to bring the parallax-based distances closer to the Sun. This effect is more extreme for the $D$(Bailer-Jones) estimator of stars that have larger spectrophotometric distances, and it is also noticeable for the BHB candidates. The overlap between the MS and BHB candidates is limited to three stars, reinforcing the validity of this method. Nevertheless, this ambiguity and the presence of seven unclassified stars suggests the use of higher-resolution spectroscopy for future analysis in addition to improved parallaxes from future {\em Gaia} releases. Finally, we note that unclassified objects may point towards the presence of either unresolved binaries or other kinds of stars, like field blue stragglers \citep{2010AJ....139...59B} that we do not account for. Note that \citet{2010AJ....139...59B} estimate that 74\% of their blue halo star sample are BHBs and 26\% are blue stragglers, which comes close to the 27 BHB stars versus 7 unclassified objects in our sample.

Table~\ref{t:spec} lists the atmospheric parameters corresponding to the spectroscopic best-fit models implied by the unique distance-driven classification as MS/BHB candidates, along with the signal-to-noise ratio (S/N) of the stacked spectra and the $E(44-55)$. In passing, we note that the 7 unclassified stars cluster around $T_{\rm eff} \approx 10\,000$\,K and $\log{g} = 3.6$. For the remainder of the manuscript, we take a conservative approach by considering as runaway MS candidates at high Galactic latitudes the 12 stars that uniquely agree within 1\,$\sigma$ of the identity line in either of the top panels of Fig.~\ref{fig:distance}, assuming the three stars with ambiguous classification  as BHB candidates.

The spectra of the two hottest stars previously mentioned (1621$-$8350 and 1944$-$4628) resemble those of MS objects of late O- and early B-type (see Fig.~\ref{fig:spectra}). However, the respective spectroscopic distances would be about 30 and 10\,kpc, which is at variance with their {\em Gaia} parallaxes (top panels of Fig.~\ref{fig:distance}). Therefore, they must be evolved low-mass stars that mimic hot massive stars. Such stars are rare but do exist, e.g., the B-type star HZ\,22 of $T_{\rm eff} = 29\,200$\,K and $\log{g} = 4.04$ \citep{saffer1997} and the O-type star PG2\,120+062 \citep{moheler1994}. HZ\,22, in fact, is a binary orbited by an unseen white dwarf, which allowed its mass to be estimated at 0.39\,M$_\sun$ \citep{schonberner1978}. Hence, those stars are likely post-RGB or post-AGB stars. For a brief account of such so-called UV-bright stars see also \citet{heber2016}.

While the SED of 1621$-$8350 can be matched by a single hot star model, this is not the case for 1944$-$4628, which shows an infrared excess indicative of a cool companion (Fig.~\ref{fig:phot-fit}). Adding a grid of SEDs from \citet{husser2013}, the infrared excess can well be matched by a composite model SED that includes a companion of $T_{\rm eff}$ = 5900$^{+1300}_{-1200}$\,K with a radius of 1.34$^{+0.60}_{-0.35}$\,R$_\odot$. Although uncertainties are large, the derived radius and luminosity of the companion are consistent with those of a G-type MS star.

We list the physical parameters of the 12 MS candidates in Table~\ref{t:ms}. The physical parameters of the 27 BHB candidates as well as those of the two post-EHB candidates and the late-type companion of 1944$-$4628 are given in Table~\ref{t:bhb_physical}.

To conclude this section, we present the results of our classification in the two theoretical Hertzsprung-Russell diagrams of Fig.~\ref{fig:hr-theo}, displaying the MS and BHB candidates with their appropriate evolutionary tracks. A few MS candidates appear to be fairly evolved, with evolutionary timescales, $\tau_{\rm evo}$, in the range of 300--400\,Myr. The BHB candidates are mostly distributed along their main locus.  From Fig.~\ref{fig:kiel}--\ref{fig:hr-theo}, we conclude that the positions of 1621$-$8350 and 1944$-$4628 are well matched by the evolution of EHB stars (subluminous B stars) after core helium exhaustion \citep{dorman1993}.
\begin{table}
\caption{Spectroscopic parameters of the sample derived from model spectra, as appropriate for their classification (MS or BHB candidate, respectively). Unclassified stars are listed at the bottom with their parameters derived from models with sub-solar composition. }
\setlength{\tabcolsep}{0.3em}
\centering
\footnotesize
\begin{tabular}{@{}lc D{o}{}{1}@{}c@{} D{,}{\,\pm\,}{2} D{,}{\,\pm\,}{3}c@{}}
\hline
Short name & S/N & \multicolumn{1}{c}{$T_{\rm eff}$} & $\log{g}$   & \multicolumn{1}{c}{$\varv_{\rm rad}$} & \multicolumn{1}{c}{$E(44-55)$} & MS \\
           &  & \multicolumn{1}{c}{(K)}  &  $\log{({\rm cm\,s}^{-2})}$ & \multicolumn{1}{c}{(km\,s$^{-1}$)} & \multicolumn{1}{c}{(mag)} & \\
\hline
0035$-$5900 & 180 & 8900 o & 3.19 & 141,31& 0.02,0.02&  \ding{51}\ding{55}\\
0037$-$6154 & 230 & 11\,900o & 4.05 & 138,26 & $<$\,0.01 & \ding{51}\\
0038$-$6612 & 220 & 8900o & 3.18 & 233,26 & $<$\,0.01 & \ding{55}\\
0039$-$4201 & 220 & 10\,300o & 3.58 & 144,25 & $<$\,0.02 & \ding{55}\\
0049$-$5356 & 240 & 9300o & 3.42 & -109,26 & $<$\,0.02 & \ding{55}\\
0122$-$6759 & 140 & 12\,800o & 3.91 & 272,27 & $<$\,0.01 & \ding{55}\\
0205$-$5938 & 380 & 12\,300o & 3.81 & 109,27 & 0.00 & \ding{51}\\
0245$-$3418 & 370 & 9300o & 3.47 & 123,28 & $<$\,0.06 & \ding{55}\\
0259$-$1705 & 370 & 12\,100o & 4.07 & 185,26 & 0.04,0.02 & \ding{51}\\
1621$-$8350$^a$ & 210 & 29\,400o & 4.00 & 115,39 & 0.08,0.02 & \ding{55}\\
1715+2341 & 190 & 12\,700o & 3.83 & 1,26 & 0.04,0.02 & \ding{55}\\
1839$-$7515 & 50 & 18\,100o & 4.53 & 65,26 & 0.12,0.03 & \ding{55}\\
1851$-$5036 & 120 & 10\,500o & 3.38 & -28,26 & 0.05,0.03 & \ding{55}\\
1900$-$7032 & 340 & 9800o & 3.58 & -46,28& 0.05,0.04&  \ding{51}\ding{55}  \\
1900$-$7033 & 250 & 16\,800o & 4.19 & -38,26 & 0.02,0.02 & \ding{51}\\
1919$-$5938 & 160 & 10\,300o & 3.61 & -88,25 & 0.06,0.04 & \ding{55}\\
1939$-$1640 & 200 & 18\,300o & 4.20 & 224,26 & 0.12,0.01 & \ding{51}\\
1944$-$4628$^{a,b}$ & 260 & 22\,300o & 4.02 & 261,26 & 0.16,0.02 & \ding{55}\\
2019$-$7107 & 320 & 10\,100o & 3.60 & 109,26& 0.05,0.03&  \ding{51}\ding{55} \\
2024$-$2725 & 180 & 10\,700o & 3.67 & 123,26 & 0.05,0.03 & \ding{55}\\
2028$-$2218 & 380 & 10\,000o & 2.84 & 79,25 & 0.05,0.03 & \ding{55}\\
2034$-$1108 & 250 & 12\,400o & 4.00 & -74,28 & 0.02,0.02 & \ding{51}\\
2034$-$5307 & 280 & 9800o & 3.61 & 123,28 & 0.04,0.03 & \ding{51}\\
2054$-$2526 & 220 & 16\,000o & 3.97 & 65,26 & 0.05,0.02 & \ding{51}\\
2110$-$2700 & 270 & 9800o & 3.48 & -3,33 & 0.06,0.03 & \ding{55}\\
2114$-$6438 & 160 & 10\,400o & 3.53 & 17,28 & 0.02,0.02 & \ding{55}\\
2119$-$4827 & 120 & 10\,400o & 3.16 & 27,25 & 0.10,0.03 & \ding{55}\\
2125$-$4834 & 230 & 10\,000o & 3.61 & 8,26 & 0.04,0.03 & \ding{55}\\
2145$-$3022 & 280 & 9700o & 3.53 & 179,27 & 0.04,0.04 & \ding{55}\\
2207$-$2645 & 340 & 12\,200o & 3.78 & 51,26 & 0.02,0.02 & \ding{51}\\
2207$-$4329 & 150 & 9700o & 3.62 & -170,26 & 0.02,0.02 & \ding{51}\\
2207$-$5021 & 190 & 9800o & 3.61 & 75,26 & 0.03,0.03 & \ding{55}\\
2208$-$6026 & 250 & 12\,000o & 3.76 & 356,25 & $<$\,0.03 & \ding{55}\\
2213$-$4357 & 250 & 12\,100o & 3.67 & 64,29 & $<$\,0.01 & \ding{55}\\
2214$-$4620 & 160 & 10\,400o & 3.62 & 112,26 & 0.01,0.01 & \ding{55}\\
2239$-$2455 & 320 & 12\,700o & 3.83 & -43,29 & $<$\,0.01 & \ding{55}\\
2250$-$5625 & 220 & 8800o & 3.20 & 118,26 & $<$\,0.01 & \ding{55}\\
2259$-$4931 & 250 & 9200o & 4.16 & 244,27 & $<$\,0.01 & \ding{51}\\
2307$-$3157 & 320 & 11\,100o & 3.94 & -2,29 & 0.01,0.01 & \ding{51}\\
2313$-$4858 & 210 & 8700o & 3.22 & -183,29 & $<$\,0.01 & \ding{55}\\
2348$-$6612 & 130 & 10000o & 3.51 & 225,26 & $<$\,0.04 & \ding{55}\\
\hline 
2040$-$2336 & 260 & 10\,200o & 3.92 & 56,26& 0.05,0.03 \\
2058$-$5407 & 300 & 9700o & 3.60 & -8,30& 0.06,0.03 \\
2103$-$5815 & 300 & 10\,000o & 3.60 & 121,27& 0.03,0.03 \\
2127$-$6300 & 220 & 9100o & 3.26 & 99,26& $<$\,0.03 \\
2304$-$4021 & 300 & 8800o & 2.99 & 27,27& 0.00 \\
2309$-$5623 & 340 & 9900o & 3.21 & 186,27& 0.00 \\
2309$-$6729 & 290 & 9800o & 3.52 & 126,32& 0.02,0.02 \\
\hline
\end{tabular}
\tablefoot{Signal-to-noise ratio, effective temperature, surface gravity, radial velocity, and interstellar reddening, and MS-candidate classification of the observed stars. Systematic uncertainties are of the order of  4\% and 0.1\,dex for $T_{\rm eff}$ and $\log{g}$, respectively. \tablefoottext{a}{This star is a hot evolved star (see Sect.~\ref{sec:spec-fitting}); based on solar-metallicity models.} \tablefoottext{b}{The result may be affected by the presence of an infrared excess (see Fig.~\ref{fig:phot-fit}).}}
\label{t:spec}
\end{table}
\begin{table}
\caption{Physical parameters of the MS candidates: mass,  radius,  luminosity, age, and spectrophotometric distance.}
\renewcommand{\arraystretch}{1.2}
\centering
\small
\begin{tabular}{@{}cccc D{o}{}{4}@{} D{o}{}{4}@{}}
\hline
Short name & $M$/M$_{\sun}$   & $R$/R$_{\sun}$   & $\log{L/{\rm L}_{\sun}}$ & \multicolumn{1}{c}{$\tau_{\rm evo}$}   & \multicolumn{1}{c}{$D$(SpT)}  \\
  	 &		  &		   &			          & \multicolumn{1}{c}{(Myr)} & \multicolumn{1}{c}{(kpc)}       \\
\hline
0037$-$6154 & $3.17^{+0.32}_{-0.17}$ & $2.77^{+0.54}_{-0.40}$ & $2.13^{+0.18}_{-0.15}$ & 213o^{+28}_{-79} & 8.34o^{+1.80}_{-1.30}\\
0205$-$5938 & $3.91^{+0.38}_{-0.29}$ & $4.07^{+0.79}_{-0.64}$ & $2.53^{+0.19}_{-0.17}$ & 152o^{+25}_{-22} & 6.19o^{+1.35}_{-1.05}\\
0259$-$1705 & $3.42^{+0.24}_{-0.21}$ & $2.83^{+0.51}_{-0.43}$ & $2.19^{+0.17}_{-0.15}$ & 148o^{+57}_{-66} & 5.04o^{+0.98}_{-0.80}\\
1900$-$7033 & $5.04^{+0.36}_{-0.25}$ & $3.00^{+0.54}_{-0.44}$ & $2.81^{+0.15}_{-0.14}$ & 38o^{+20}_{-24} & 8.19o^{+1.57}_{-1.25}\\
1939$-$1640 & $5.81^{+0.36}_{-0.35}$ & $3.16^{+0.55}_{-0.47}$ & $3.00^{+0.15}_{-0.15}$ & 27o^{+16}_{-20} & 4.72o^{+0.86}_{-0.73}\\
2034$-$1108 & $3.58^{+0.28}_{-0.24}$ & $3.14^{+0.58}_{-0.48}$ & $2.32^{+0.17}_{-0.16}$ & 165o^{+26}_{-36} & 3.48o^{+0.68}_{-0.56}\\
2034$-$5307 & $3.00^{+0.37}_{-0.11}$ & $4.51^{+0.94}_{-0.60}$ & $2.23^{+0.19}_{-0.14}$ & 317o^{+48}_{-59} & 6.01o^{+1.34}_{-0.85}\\
2054$-$2526 & $5.21^{+0.48}_{-0.36}$ & $3.91^{+0.74}_{-0.60}$ & $2.95^{+0.17}_{-0.15}$ & 63o^{+9}_{-13} & 8.22o^{+1.68}_{-1.33}\\
2207$-$2645 & $3.90^{+0.26}_{-0.21}$ & $4.20^{+0.75}_{-0.62}$ & $2.54^{+0.15}_{-0.14}$ & 171o^{+23}_{-21} & 3.96o^{+0.75}_{-0.61}\\
2207$-$4329 & $2.97^{+0.31}_{-0.12}$ & $4.45^{+0.88}_{-0.60}$ & $2.21^{+0.17}_{-0.14}$ & 329o^{+50}_{-58} & 5.71o^{+1.21}_{-0.81}\\
2259$-$4931 & $2.17^{+0.18}_{-0.20}$ & $2.03^{+0.38}_{-0.33}$ & $1.43^{+0.16}_{-0.21}$ & 462o^{+160}_{-212} & 3.97o^{+0.80}_{-0.79}\\
2307$-$3157 & $3.15^{+0.29}_{-0.24}$ & $3.14^{+0.60}_{-0.49}$ & $2.13^{+0.17}_{-0.16}$ & 236o^{+26}_{-36} & 3.01o^{+0.61}_{-0.49}\\
\hline
\label{t:ms}
\end{tabular}
\end{table}
\begin{table}
\caption{Physical parameters of BHB and post-EHB candidates computed adopting the atmospheric parameters derived  from metal deficient atmospheric models and a mass of $M = 0.5 \pm 0.1$\,M$_{\sun}$. The two hot post-EHB stars are listed at the end. 1944$-$4628B denotes the cool companion of 1944$-$4628.} 
\small
\centering
\begin{tabular}{@{}lcc@{} D{o}{}{4}@{}}
\hline
Short name & $R$/R$_{\sun}$   & $\log{L/{\rm L}_{\sun}}$ & \multicolumn{1}{c}{$D$(SpT)}\\
  	 &		  &		   &		\multicolumn{1}{c}{(kpc)}  \\
\hline
0035$-$5900 & $2.94^{+0.48}_{-0.44}$ & $1.68^{+0.15}_{-0.16}$ & 5.84o^{+1.03}_{-0.92}\\
0038$-$6612 & $2.97^{+0.48}_{-0.44}$ & $1.70^{+0.15}_{-0.16}$ & 4.54o^{+0.78}_{-0.70}\\
0039$-$4201 & $1.88^{+0.31}_{-0.29}$ & $1.56^{+0.15}_{-0.16}$ & 3.61o^{+0.62}_{-0.55}\\
0049$-$5356 & $2.28^{+0.37}_{-0.34}$ & $1.54^{+0.15}_{-0.16}$ & 3.48o^{+0.58}_{-0.53}\\
0122$-$6759 & $1.29^{+0.21}_{-0.20}$ & $1.60^{+0.15}_{-0.16}$ & 4.15o^{+0.70}_{-0.64}\\
0245$-$3418 & $2.15^{+0.36}_{-0.32}$ & $1.50^{+0.15}_{-0.16}$ & 1.70o^{+0.28}_{-0.26}\\
1715+2341 & $1.41^{+0.23}_{-0.21}$ & $1.67^{+0.15}_{-0.16}$ & 3.51o^{+0.59}_{-0.54}\\
1839$-$7515 & $0.63^{+0.12}_{-0.10}$ & $1.59^{+0.17}_{-0.17}$ & 4.27o^{+0.79}_{-0.68}\\
1851$-$5036 & $2.37^{+0.38}_{-0.35}$ & $1.78^{+0.15}_{-0.16}$ & 6.82o^{+1.12}_{-1.01}\\
1900$-$7032 & $1.88^{+0.31}_{-0.28}$ & $1.47^{+0.15}_{-0.16}$ & 2.35o^{+0.40}_{-0.35}\\
1919$-$5938 & $1.83^{+0.30}_{-0.27}$ & $1.52^{+0.15}_{-0.16}$ & 3.87o^{+0.64}_{-0.58}\\
2019$-$7107 & $1.84^{+0.30}_{-0.28}$ & $1.50^{+0.15}_{-0.16}$ & 1.83o^{+0.30}_{-0.28}\\
2024$-$2725 & $1.71^{+0.28}_{-0.26}$ & $1.54^{+0.15}_{-0.16}$ & 3.16o^{+0.52}_{-0.49}\\
2028$-$2218 & $4.41^{+0.71}_{-0.66}$ & $2.24^{+0.15}_{-0.16}$ & 3.61o^{+0.59}_{-0.55}\\
2110$-$2700 & $2.13^{+0.35}_{-0.31}$ & $1.57^{+0.15}_{-0.16}$ & 2.56o^{+0.43}_{-0.38}\\
2114$-$6438 & $1.99^{+0.34}_{-0.30}$ & $1.62^{+0.15}_{-0.16}$ & 3.69o^{+0.63}_{-0.56}\\
2119$-$4827 & $3.04^{+0.53}_{-0.45}$ & $1.99^{+0.16}_{-0.16}$ & 6.51o^{+1.14}_{-0.98}\\
2125$-$4834 & $1.84^{+0.31}_{-0.28}$ & $1.48^{+0.15}_{-0.16}$ & 3.38o^{+0.56}_{-0.51}\\
2145$-$3022 & $2.01^{+0.33}_{-0.30}$ & $1.50^{+0.15}_{-0.16}$ & 2.91o^{+0.49}_{-0.44}\\
2207$-$5021 & $1.83^{+0.31}_{-0.28}$ & $1.45^{+0.15}_{-0.16}$ & 2.76o^{+0.47}_{-0.42}\\
2208$-$6026 & $1.54^{+0.25}_{-0.23}$ & $1.64^{+0.15}_{-0.16}$ & 3.06o^{+0.52}_{-0.47}\\
2213$-$4357 & $1.70^{+0.28}_{-0.25}$ & $1.75^{+0.15}_{-0.16}$ & 3.29o^{+0.55}_{-0.50}\\
2214$-$4620 & $1.81^{+0.30}_{-0.27}$ & $1.53^{+0.15}_{-0.16}$ & 4.89o^{+0.83}_{-0.74}\\
2239$-$2455 & $1.41^{+0.23}_{-0.21}$ & $1.66^{+0.15}_{-0.16}$ & 1.64o^{+0.27}_{-0.24}\\
2250$-$5625 & $2.91^{+0.48}_{-0.44}$ & $1.65^{+0.15}_{-0.16}$ & 5.58o^{+0.99}_{-0.90}\\
2313$-$4858 & $2.87^{+0.48}_{-0.43}$ & $1.63^{+0.15}_{-0.16}$ & 4.61o^{+0.84}_{-0.73}\\
2348$-$6612 & $2.04^{+0.35}_{-0.32}$ & $1.56^{+0.16}_{-0.17}$ & 7.02o^{+1.22}_{-1.11}\\
\hline
1621$-$8350$^{a}$ & $1.16^{+0.20}_{-0.18}$ & $3.09^{+0.18}_{-0.19}$ & 5.22o^{+0.92}_{-0.82}\\
1944$-$4628$^{a}$ & $1.14^{+0.19}_{0.17}$ & 2.46 $^{+0.16}_{-0.16}$ & 2.56o^{+0.44}_{-0.38}\\
1944$-$4628B$^{a}$ & 1.34$^{+0.60}_{-0.35}$ & -0.06 $^{+0.68}_{-0.90}$ & \\
\hline
\end{tabular}
\label{t:bhb_physical}
\tablefoot{\tablefoottext{a}{Based on solar-metallicity models.}}
\end{table}

\section{Galactic orbit simulation}
\label{sec:orbits}
To model the trajectories of our  MS candidates in the Milky Way's gravitational potential, we use the {\sc python} package {\sc galpy} for galactic dynamics \citep{bovy2015}. We adopt the Model\,I potential of \citet{irrgang2013}, which is implemented in {\sc galpy} in a simplified fashion that serves our purposes. This potential is a revised version of that described by \citet{allen-santillan1991}. The Sun is located at $R_{\sun} = 8.4$\,kpc from the Galactic center, around which it orbits with $\varv_{0} = 242$\,km\,s$^{-1}$. The local escape velocity is  $\varv_{\sun,\,{\rm esc}} = 614.4$\,km\,s$^{-1}$. We note that the {\sc galpy} reference potential, \texttt{MWPotential2014}, accounts for a lower-mass Galactic halo model that, however, has relatively small impact in determining the past trajectories of the youngest and nearest runaway candidates in our sample. The mass of the Milky Way's halo can be relevant for the study of unbound stars, as we discuss in the following section.

The past and future trajectories of the runaway candidates are determined in a statistical fashion, as we randomly sample the observed quantities ($\alpha$, $\delta$, $\mu^{*}_{\alpha}$, $\mu_\delta$, $D(\rm SpT)$, $\varv_{\rm rad}$) by adopting their uncertainties and relevant correlations. We draw 10\,000 initial conditions, which are evolved for 10\,Gyr in the future and traced back in the past for sufficiently large time (i.e.\ up to $\tau_{\rm evo}$). For each star, we note the rest-frame velocity, $\varv_{\rm rf}$, the past and future Galactic cartesian coordinates, and the corresponding velocity components. We estimate the escape probability from the Milky Way, $P_{\rm esc}$, for each star as the fraction of initial conditions where $(\varv_{\rm rf} - \varv_{\rm esc}) > 0$. We determine the vertical component  of the angular momentum, $L_Z$, which indicates the direction towards which the star has been ejected. We also numerically estimate the eccentricity, $e$, the median orbital apocenter and pericenter, $R_{\rm apo}$ and $R_{\rm peri}$, and the maximum vertical displacement, $Z_{\rm max}$. The most recent Galactic plane crossings, $R(Z = 0)$, are recorded and used to determine the flight time from such locations, $\tau_{\rm flight}$. The ejection velocity, $V_{\rm ej}$, is determined as the velocity possessed by the star at $R(Z=0)$, corrected for the solar motion and the Galactic rotation.

Table~\ref{t:orbit} lists the relevant orbital quantities for the 12 MS runaway candidates in a Galactocentric reference system that is left-handed, i.e., with the $x$- and $y$-axis pointing, respectively, from the Milky Way's center towards the Sun and along the direction of the Galactic rotation.

\begin{table*}
\caption{Orbital parameters of our MS candidates.}
\small
\centering
\setlength{\tabcolsep}{0.101cm}
\renewcommand{\arraystretch}{1.2}
\begin{tabular}{@{}c D{o}{}{4}@{} D{o}{}{4}@{} D{o}{}{4}@{} D{o}{}{4}@{} D{o}{}{4}@{} cc@{} D{o}{}{4}@{}  D{o}{}{4}@{} D{o}{}{4}@{} D{o}{}{4}@{} D{o}{}{5}@{} c@{} D{o}{}{3}@{}}
\hline
Short name & \multicolumn{1}{c}{$X$}   & \multicolumn{1}{c}{$Y$}	& \multicolumn{1}{c}{$Z$}   & \multicolumn{1}{c}{$\varv_{Z}$}  & \multicolumn{1}{c}{$\varv_{\rm rf}$}  & $P_{\rm esc}$  & $e$ & \multicolumn{1}{c}{$L_{Z}$}           & \multicolumn{1}{c}{$R_{\rm peri}$}      & \multicolumn{1}{c}{$R_{\rm apo}$} & \multicolumn{1}{c}{$Z_{\rm max}$} & \multicolumn{1}{c}{$R(Z=0)$} & $\varv_{\rm ej}$   & \multicolumn{1}{c}{$\tau_{\rm flight}$} \\
	       & \multicolumn{1}{c}{(kpc)} & \multicolumn{1}{c}{(kpc)} & \multicolumn{1}{c}{(kpc)} & \multicolumn{1}{c}{(km\,s$^{-1}$)}& \multicolumn{1}{c}{(km\,s$^{-1}$)} &	         &     & \multicolumn{1}{c}{(kpc\,km\,s$^{-1}$)}& \multicolumn{1}{c}{(kpc)}   & \multicolumn{1}{c}{(kpc)}     & \multicolumn{1}{c}{(kpc)}          & \multicolumn{1}{c}{(kpc)}    & (km\,s$^{-1}$) & \multicolumn{1}{c}{(Myr)}		     \\
\hline
0037$-$6154 & 5.63o^{+0.50}_{-0.53} & -3.86o^{+0.70}_{-0.74} & -6.82o^{+1.24}_{-1.31} & -231o^{+31}_{-32} & 380o^{+42}_{-38} & $0.0$ & $0.7$ & 2100o & 8.24o^{+0.91}_{-0.62} & 40.6o^{+29.0}_{-15.0}  & 31.5o^{+25.0}_{-12.7} & 8.9o^{+1.6}_{-1.1}  & $330^{+31}_{-37}$ & 24o^{+3}_{-3} \\
0205$-$5938 & 7.40o^{+0.19}_{-0.20} & -3.41o^{+0.66}_{-0.67} & -5.07o^{+0.98}_{-1.00} & -61o^{+22}_{-23} & 194o^{+18}_{-12} & $0.0$ & $0.4$ & 1300o & 4.81o^{+0.88}_{-0.66} & 11.49o^{+1.86}_{-1.36}  & 6.17o^{+1.71}_{-1.28} & 5.2o^{+0.7}_{-0.5}  & $273^{+84}_{-47}$ & 37o^{+6}_{-6} \\
0259$-$1705 & 10.85o^{+0.43}_{-0.43} & -0.93o^{+0.16}_{-0.16} & -4.30o^{+0.76}_{-0.76} & -178o^{+22}_{-23} & 204o^{+18}_{-16} & $0.0$ & $0.2$ & 1000o & 7.90o^{+0.75}_{-0.66} & 12.65o^{+1.10}_{-0.83}  & 11.22o^{+1.64}_{-1.69} & 10.5o^{+0.7}_{-0.6}  & $276^{+27}_{-27}$ & 21o^{+4}_{-4} \\
1900$-$7033 & 2.43o^{+1.01}_{-1.03} & -4.23o^{+0.72}_{-0.73} & -3.60o^{+0.61}_{-0.63} & -183o^{+37}_{-38} & 363o^{+30}_{-28} & $0.0$ & $0.5$ & 1300o & 5.77o^{+0.60}_{-0.23} & 19.41o^{+9.36}_{-5.21}  & 15.20o^{+9.70}_{-5.72} & 7.9o^{+1.0}_{-0.8}  & $295^{+25}_{-39}$ & 16o^{+1}_{-1} \\
1939$-$1640 & 4.27o^{+0.69}_{-0.69} & 1.75o^{+0.29}_{-0.29} & -1.44o^{+0.24}_{-0.25} & -166o^{+19}_{-19} & 340o^{+25}_{-25} & $0.0$ & $0.5$ & 1100o & 4.12o^{+0.19}_{-0.11} & 11.79o^{+3.10}_{-2.33}  & 6.77o^{+1.60}_{-1.23} & 5.9o^{+0.6}_{-0.6}  & $259^{+23}_{-23}$ & 8o^{+0}_{-1} \\
2034$-$1108 & 5.87o^{+0.45}_{-0.45} & 1.74o^{+0.31}_{-0.31} & -1.62o^{+0.29}_{-0.29} & 188o^{+29}_{-29} & 253o^{+28}_{-24} & $0.0$ & $0.4$ & 900o & 3.64o^{+0.31}_{-0.27} & 9.15o^{+1.63}_{-1.06}  & 6.19o^{+1.62}_{-1.30} & 3.8o^{+0.4}_{-0.3}  & $409^{+30}_{-42}$ & 93o^{+19}_{-13} \\
2034$-$5307 & 3.75o^{+0.83}_{-0.83} & -1.22o^{+0.22}_{-0.22} & -3.57o^{+0.64}_{-0.64} & 102o^{+35}_{-36} & 339o^{+96}_{-94} & $0.0$ & $0.5$ & -1200o & 4.89o^{+0.06}_{-0.45} & 14.0o^{+21.0}_{-6.6}  & 9.1o^{+18.4}_{-5.2} & 13.2o^{+15.0}_{-6.3}  & $371^{+63}_{-48}$ & 68o^{+45}_{-19} \\
2054$-$2526 & 2.30o^{+1.09}_{-1.12} & 2.28o^{+0.42}_{-0.41} & -4.98o^{+0.89}_{-0.92} & 48o^{+22}_{-22} & 221o^{+25}_{-21} & $0.0$ & $0.3$ & 700o & 3.85o^{+1.15}_{-0.68} & 7.75o^{+1.51}_{-0.75}  & 6.48o^{+1.73}_{-1.37} & 6.9o^{+1.7}_{-1.2}  & $262^{+16}_{-17}$ & 50o^{+11}_{-7} \\
2207$-$2645 & 6.26o^{+0.36}_{-0.37} & 0.95o^{+0.16}_{-0.16} & -3.17o^{+0.54}_{-0.55} & -152o^{+28}_{-29} & 342o^{+25}_{-23} & $0.0$ & $0.6$ & 1600o & 4.41o^{+0.72}_{-0.66} & 20.36o^{+5.25}_{-3.63}  & 9.56o^{+3.61}_{-2.71} & 4.4o^{+0.7}_{-0.6}  & $272^{+47}_{-44}$ & 16o^{+1}_{-1} \\
2207$-$4329 & 5.00o^{+0.60}_{-0.60} & -0.26o^{+0.05}_{-0.05} & -4.57o^{+0.81}_{-0.81} & -50o^{+41}_{-40} & 638o^{+140}_{-138} & $0.4$ & ... & -2300o & 6.78o^{+0.18}_{-2.06} & ...  & ... & 16.5o^{+5.9}_{-5.4}  & $802^{+129}_{-115}$ & 32o^{+4}_{-3} \\
2259$-$4931 & 6.50o^{+0.37}_{-0.38} & -0.73o^{+0.14}_{-0.15} & -3.38o^{+0.67}_{-0.68} & -114o^{+28}_{-30} & 350o^{+93}_{-82} & $0.0$ & $0.5$ & -2100o & 7.25o^{+0.03}_{-0.06} & 20.9o^{+33.5}_{-10.8}  & 11.2o^{+19.0}_{-4.9} & 9.0o^{+2.8}_{-1.0}  & $541^{+40}_{-54}$ & 22o^{+6}_{-5} \\
2307$-$3157 & 7.26o^{+0.21}_{-0.21} & 0.30o^{+0.05}_{-0.06} & -2.77o^{+0.51}_{-0.51} & -101o^{+34}_{-33} & 230o^{+40}_{-32} & $0.0$ & $0.9$ & 200o & 0.63o^{+0.69}_{-0.48} & 12.92o^{+3.13}_{-1.93}  & 6.17o^{+5.33}_{-2.35} & 2.0o^{+1.4}_{-1.2}  & $469^{+170}_{-106}$ & 18o^{+2}_{-2} \\

\hline
\end{tabular}
\label{t:orbit}
\tablefoot{The meaning of the columns from left to right is: current $XYZ$ Galactocentric Cartesian coordinates, vertical component of Galactocentric rest-frame velocity, Galactocentric rest-frame velocity, escape  probability, orbit eccentricity, vertical component of the angular momentum, median pericenter and apocenter, maximum vertical displacement from the plane, Galactocentric radius at the most recent intersection of past trajectories with the $Z=0$ plane, ejection velocity from $Z = 0$, and flight time from $Z = 0$.}
\end{table*}
\section{Properties of the runaway candidates}
\label{sec:discussion}
\subsection{Orbit analysis}
As expected for nearby runaway MS stars, all but one candidate are bound to the Milky Way's potential, since they have $\varv_{\rm rf}$ that is on average two times smaller than $\varv_{\rm esc}$ at their Galactic coordinates. The  only exceptions is 2207$-$4329 that has an escape probability $P_{\rm esc} = 0.4$, defined as the fraction of simulated trajectories with $\varv_{\rm rest} > \varv_{\rm esc}$.

\begin{figure}
   \centering
   \includegraphics[width=\linewidth]{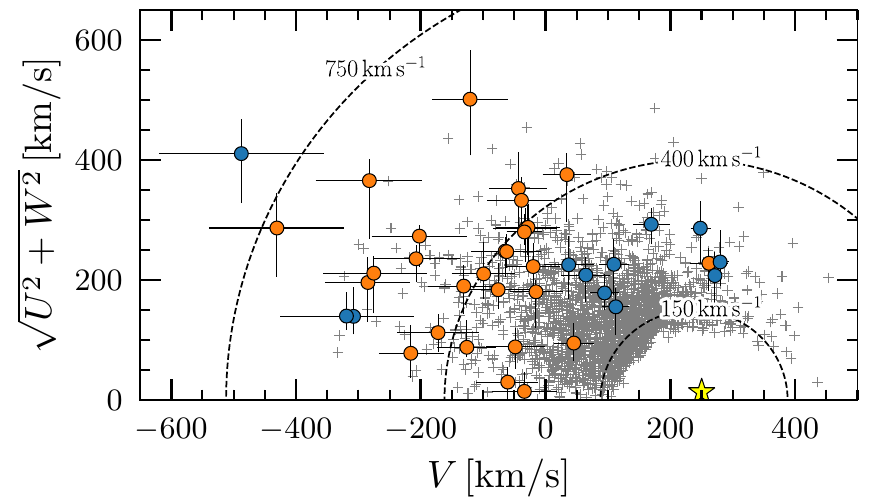}
   \caption{The positions of the program stars in the Toomre diagram. The quantity $V$ is the velocity component in direction of Galactic rotation, $U$ towards the Galactic center, and $W$ perpendicular to the Galactic plane.  The local standard of rest (LSR) is marked by a yellow asterisk. The dashed circles centered around the LSR represent boundaries for three restframe velocities. The observed stars are plotted with error bars. Blue symbols correspond to MS candidates, while orange symbols are used for the BHB stars. All the targets identified in Sect.~\ref{sec:selection} are plotted as gray crosses, adopting $\varv_{\rm rad} = 0$\,km\,s$^{-1}$.}
    \label{fig:toomre}
    \end{figure}

 Due to our selection cut of $\varv_{\rm tan} > 150$\,km\,s$^{-1}$ (Eq.~\ref{eq:vtan}), all stars in our sample (both the MS and BHB candidates) resulted to have $|\varv_{\rm rad}| \lesssim \varv_{\rm tan}$ within the error bars. Three stars, 2034$-$5307, 2207$-$4329, and 2259$-$4931, have the largest $\varv_{\rm tan} \gtrsim 450$\,km\,s$^{-1}$ of all MS candidates. This unusual feature can be also noted in the Toomre diagram of Fig.~\ref{fig:toomre}, where the three stars have large negative Galactic cartesian $V$ components (i.e. they lag behind the Milky Way's rotation), as it is more typically seen for the majority of BHB candidates in our sample. The remaining MS candidates, instead, have moderate tangential velocities of $\varv_{\rm tan} \simeq 200$\,km\,s$^{-1}$. 
The mostly retrograde motions of the BHB stars implies that they belong to the halo population. In the future, additional runaway candidates could be selected on the basis of their location in the Galactic Cartesian velocity space, as depicted by the Toomre diagram (Fig.~\ref{fig:toomre}).  
 
 Following the general direction of the Galactic rotation most MS candidates have $L_Z > 0$\,kpc\,km\,s$^{-1}$, with the exception of the three more extreme stars that have large negative, Cartesian velocity $V$ components. While our distance-based method entirely classify them as MS stars, these three stars appear to possess kinematics that are more typical of BHB stars. We note that 2034$-$5307 and 2207$-$4329 have $T_{\rm eff}$ and $\log{g}$ in the overlapping region of MS tracks and the horizontal branch (Fig.~\ref{fig:kiel}). Instead, 2259$-$4931 has a larger $\log{g}$ that excludes a BHB nature. As already mentioned in Sect.~\ref{sec:analysis}, our distance-based method identifies MS candidates for which we envisage higher-resolution spectroscopy for a more detailed assessment of their nature.

All the MS candidates are found below the Galactic plane and are currently moving away from it, with the exception of three stars, 2034$-$1108, 2034$-$5307, and 2054$-$2526, that are approaching the plane. The orbits of the observed runaway candidates cover a wide range of eccentricities, $e = 0.2$--0.9, that is larger than the typical values for thin disc stars. All the runaway candidates have their orbital apocenters roughly between the solar circle and $R_{\rm apo} = 20$\,kpc, with a few exceptions potentially reaching out to very far Galactic radii. Their orbits are typically enclosed within  $Z_{\rm max} = 4$--40\,kpc from the Galactic plane.

The trajectories of 2307$-$3157 have their orbital pericenters relatively close to the Galactic center ($R_{\rm peri} < 1$\,kpc); hence, its motion could be affected by the shape and orientation of the Galactic bar, which is not accounted for in our Galactic  potential model. The trajectories of stars reaching beyond $R_{\rm apo} \gtrsim 20$\,kpc, instead, might experience gravitational perturbations due to the Large Magellanic Cloud \citep{kenyon2018,erkal2019}, which our Galactic model also does not account for. 

Finally, we do not list the orbital parameters ($e$, $R_{\rm peri}$, $R_{\rm apo}$, and $Z_{\rm max}$) of 2207$-$4329 in Table~\ref{t:orbit}, for they are unconstrained due to its  non-zero probability of escaping the Milky Way. For this star, a significant fraction of simulated trajectories reaches extremely large Galatocentric  radii without returning back, within the 10\,Gyr time-frame of our simulation. As for the other stars, the parallax uncertainty has a major role in determining the agreement with the spectrophotometric distance and its classification as MS candidate. For the time being, we note that the adopted potential \citep[Model\,I of][]{irrgang2013} delivers a $\varv_{\sun,\,{\rm esc}}$ that is already a few tens of km\,s$^{-1}$ larger than more recent {\em Gaia}-based estimates \citep{monari2018,deasons2019}, suggesting a real chance for this star to escape the Milky Way if confirmed as MS candidate.

\subsection{Birth  place and flight times}

The birth places of B-type MS stars are open clusters and associations within the thin disc of the Milky Way. Although short compared to that of lower mass stars, the MS life of late-B and early-A type stars can span from several 10\,Myr up to $\sim 1$\,Gyr, making the identification of their birth places a relatively complex issue. 

In Sect.~\ref{sec:orbits}, we have determined the most recent crossings of the Galactic plane for the MS candidates, from which we estimate the flight times. To confirm the runaway hypothesis, the flight times need to be shorter than the ages estimated from evolutionary tracks ($\tau_{\rm flight} \leq \tau_{\rm evo}$). In the case of  ejection via the binary supernova mechanism, a runaway star could be rejuvenated due to mass-transfer prior to the core-collapse explosion of the primary \citep{schneider2015}; hence, it could even appear slightly younger than the flight time.

\begin{figure}
   \centering
   \includegraphics[width=\linewidth]{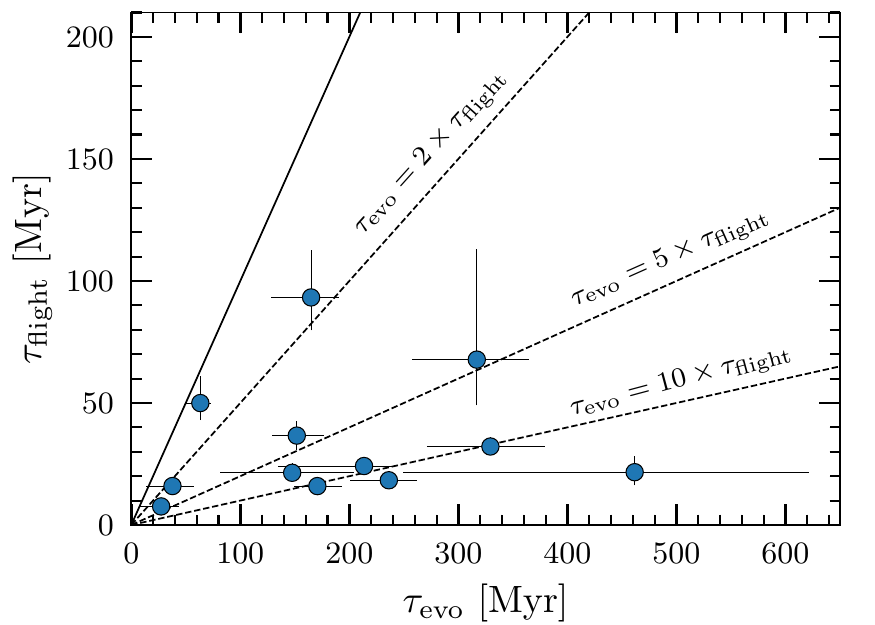}
   \caption{Comparison of flight times and stellar ages for the runaway MS candidates. The solid  line represents the identity curve. The dashed lines mark the ranges where $\tau_{\rm evo} = 2,\,5$, and 10 times $\tau_{\rm flight}$.}
    \label{fig:age-flight}
    \end{figure}

In Fig.~\ref{fig:age-flight}, we show the comparison between these two quantities, noting that -- within the error bars -- all the candidates are below or overlap with the solid curve representing the identity line. Two thirds of our runaway candidates have ages that are more than 2 times longer than the flight times. This result poses the question of whether the reported crossing locations of the Galactic plane, $R(Z=0)$, identify the real birth places and flight times. In fact, we note that all candidates with $\tau_{\rm evo} > 100$\,Myr, with the exclusion of 2207$-$4329, could have crossed the Galactic plane two or more times in the past.  If the mechanisms that caused their runaway status occurred early in their lifetime, as it is generally accepted for the binary supernova and dynamical ejection scenarios, the most recent $R(Z=0)$ do not constrain their $\tau_{\rm flight}$.

The various ejection mechanisms have intrinsic timescales.   For example, the canonical binary ejection mechanism necessarily occurs on very short timescales of a few Myr, which are driven by the  core-collapse supernova \citep{portegieszwart2000}, and the dynamical ejection from clusters can produce runaway stars for several hundred Myr \citep{moyanoloyola2013}. Therefore, future intermediate- and high-resolution spectroscopy would enable an improved understanding, e.g. via the detection of polluting elements from supernova ejecta \citep{pan2012} as found for the hyper-runaway giant \object{HD\,271971} \citep{przybilla2008} or via the comparison of measured abundances with the average elemental abundances of specific locations in the Galactic disc \citep{irrgang2010}.

Of all MS candidates, the possibly unbound runaway, 2207$-$4329, also has the Galactic plane crossing radius that is the furthest from the Galactic center at $R(Z=0) = 16.5^{+5.9}_{-5.4}$. Such a large distance is similar to the case of HD\,271791 also originating from the outer Galactic disk \citep{2008A&A...483L..21H,przybilla2008}, the ejection velocity of which, however, is much lower \citep[$390\pm4$\,km\,s$^{-1}$;][]{irrgang2019}. If the MS classification of 2207$-$4329 is confirmed by future {\em Gaia} data releases and higher-resolution spectroscopy, alternative birth places above the Galactic plane \citep[e.g.\ the tidal stream of a satellite galaxy colliding with the Milky Way;][]{abadi2009} or other exotic scenarios involving intermediate-mass black holes \citep{fragione2019} should also be investigated.

\subsection{Ejection velocity distribution}
   \begin{figure}
   \centering
   \includegraphics[width=\linewidth]{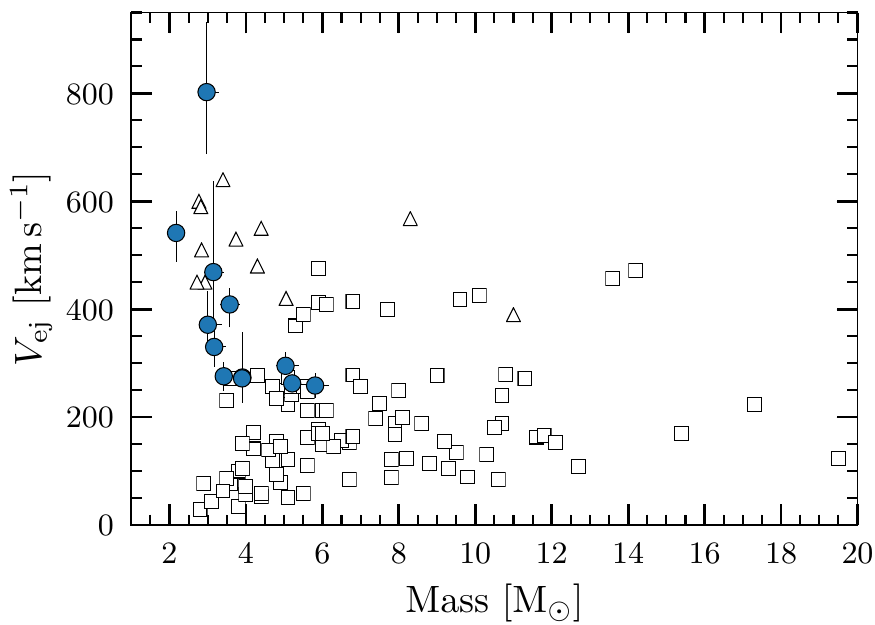}
   \caption{Ejection velocities of
   runaway stars are plotted against their masses. Our 12 MS runaway candidates are represented by blue circles with error bars (note that two stars have very similar masses and, thus, overlapping symbols). The runaway stars studied by \citet{silva2011} are plotted  as squares. More recently discovered high-velocity runaway stars as listed by \citet{irrgang2019} are plotted as triangles, with masses updated from \citet{kreuzer2020} when available.}
    \label{fig:vej-mass}
    \end{figure}

\citet{silva2011} described the relation between stellar mass and ejection velocity, with reference to their compiled sample of runaway stars, as mostly determined by the binary supernova mechanism. They also noted that the dynamical ejection from clusters could contribute to the low-ejection velocity population. On the contrary, the fastest runaway stars could have an overlap with the slowest hyper-velocity stars.

In Fig.~\ref{fig:vej-mass}, our runaway candidates occupy a lower mass range with respect to the \citet{silva2011} and \citet{irrgang2019} samples. The increase of ejection velocities towards lower masses that characterize our sample appears to be in agreement with the predictions for binary supernova ejections in the most extreme scenarios \citep{tauris2015, evans2020}. However, \citet{renzo2019} suggest that the dominant fraction of runaway stars with large ejection velocities could be mostly formed via dynamical interactions in open clusters, or more exotic scenarios.
Interestingly, the ejection velocities of three runaway candidates, 2207$-$4329, 2259$-$4931, and 2307$-$3157, could largely exceed $400$--$500$\,km\,s$^{-1}$, which would be challenging the classical runaway scenarios \citep{irrgang2018b,irrgang2019}. Of the three stars, 2207$-$4329 is by far the fastest (802$^{+129}_{-115}$\,km\,s$^{-1}$). 
2259$-$4931, which is the least massive star of our sample, is the second fastest one, ejected at 541$^{+40}_{-54}$\,km\,s$^{-1}$ from near the solar circle. If confirmed as a MS star, 2207$-$4329 would be the fastest known hyper-runaway star \citep[cf. Table 5 of][]{irrgang2019}. Its ejection velocity would exceed that of HVS\,5 \citep[640$^{+50}_{-40}$\,km\,s$^{-1}$;][]{irrgang2019}, even when we allow for the rather large uncertainty of the former. 2259$-$4931 would also gain a prominent position in the list of Galactic disc runaways with the largest ejection velocities, next to the well studied PG\,1610+062 (550$\pm$20\,km\,s$^{-1}$).

\section{Conclusions}

We have presented the selection of bright early-type MS candidates at high Galactic latitudes,  relying on the {\em Gaia} DR2 astrometry and photometry, with the goal of identifying new runaway stars. To test our selection we have followed-up 48 stars obtaining low-resolution optical spectroscopy. By comparing spectrophotometric distances and parallax-based distance estimators, we propose 12 stars as MS candidates. Of the remaining stars, 27 are likely BHB stars, two are rare highly evolved hot stars, and  seven yet unclassified stars could be binaries or blue stragglers. 

The newly identified runaway candidates are mostly found within the 2--4\,M$_\sun$ range. We numerically trace the past trajectories of these stars, identifying their nearest Galactic disc crossings locations, flight times, and ejection velocities. Most stars have ejection velocities between 200--450\,km\,s$^{-1}$. Three stars exceed the 450\,km\,s$^{-1}$ upper limit of classical ejection scenarios \citep[see][]{irrgang2019} such as the binary supernova mechanism and dynamical ejection from open clusters and, thus, call for a yet unknown ejection process \citep[see also][]{evans2020}. One of the three stars has a non-negligible escape probability, but more precise and accurate astrometry is required to check whether they are bound to the Galaxy. According to their kinematics the BHB stars belong to the halo population.

We remark that high-resolution spectroscopy is crucial to asses precise rotational velocities and abundances, which are necessary to confirm the MS nature of these candidates and shed light on their ejection mechanisms and birth places. We also envisage an improved kinematic analysis and stellar classification of the wider target lists, arising from the future {\em Gaia} data releases. More accurate and precise parallaxes will make it possible to extend the current volume beyond $\sim$10\,kpc, which is currently limited by the parallax precision and accuracy.

\begin{acknowledgements}
RR, AI, and UH acknowledge funding by the German Science foundation (DFG) through grants HE1356/71-1 and IR190/1-1. DS was supported by the Deutsche Forschungsgemeinschaft (DFG) under grants HE1356/70-1 and IR190/1-1. RR has received funding from the postdoctoral fellowship programme Beatriu de Pin\'os, funded by the Secretary of Universities and Research (Government of Catalonia) and by the Horizon 2020 programme of research and innovation of the European Union under the Maria Sk\l{}odowska-Curie grant agreement No 801370.

We thank I. Condor for his support during the NTT observing run.

This work has made use of data from the European Space Agency (ESA)
mission {\it Gaia} (\url{https://www.cosmos.esa.int/gaia}), processed by
the {\it Gaia} Data Processing and Analysis Consortium (DPAC,
\url{https://www.cosmos.esa.int/web/gaia/dpac/consortium}). Funding
for the DPAC has been provided by national institutions, in particular
the institutions participating in the {\it Gaia} Multilateral Agreement. Based on observations collected at the European Organisation for Astronomical Research in the Southern Hemisphere under ESO programme 0103.D-0530. 

This research has made use of the VizieR catalogue access tool, CDS, Strasbourg, France (DOI : 10.26093/cds/vizier). The original description 
 of the VizieR service was published in 2000, A\&AS 143, 23
 This work made use of the IPython package \citep{PER-GRA:2007} This publication makes use of data products from the Two Micron All Sky Survey, which is a joint project of the University of Massachusetts and the Infrared Processing and Analysis Center/California Institute of Technology, funded by the National Aeronautics and Space Administration and the National Science Foundation. Based on observations made with the NASA Galaxy Evolution Explorer. GALEX is operated for NASA by the California Institute of Technology under NASA contract NAS5-98034.  This publication makes use of data products from the Wide-field Infrared Survey Explorer \citep{wright2010}, which is a joint project of the University of California, Los Angeles, and the Jet Propulsion Laboratory/California Institute of Technology, funded by the National Aeronautics and Space Administration. 

\end{acknowledgements}

\appendix

\section{Gaia archive query}
\label{a:one}

We performed the following query via the online service
offered by the Astronomisches Rechen-Institut of the
Heidelberg University\footnote{\url{http://gaia.ari.uni-heidelberg.de/}}:

\begin{verbatim}
SELECT * 
FROM gaiadr2.gaia_source 
WHERE parallax_error/parallax<=0.3 
   AND parallax  > 0 
   AND phot_g_mean_mag + 5 - 
       5*log10(1000/parallax)<= 3.7 
   AND ABS(b)>=15 AND bp_rp<=0.05
   AND phot_bp_rp_excess_factor 
       <= 1.3 + 0.06*power(bp_rp,2)
   AND phot_bp_rp_excess_factor 
       >= 1.0 + 0.015*power(bp_rp,2)
   AND visibility_periods_used >= 8
   AND phot_g_mean_flux_over_error >= 50
   AND phot_bp_mean_flux_over_error >= 20
   AND phot_rp_mean_flux_over_error >= 20
   AND ruwe <= 1.4
\end{verbatim}
In addition, we filtered the data with the prescription of
Eq.\ref{eq:vtan}.

%
%

\bibliographystyle{aa}
\bibliography{bibliography}
\end{document}